\documentclass[prx,longbibliography,twocolumn,notitlepage,showpacs,amsmath,amstex,amssymb,citeautoscript,superscriptaddress]{revtex4-1}

\usepackage[english]{babel}
\usepackage{letltxmacro}
\usepackage{latexsym}
\usepackage[shortlabels]{enumitem}

\LetLtxMacro{\ORIGselectlanguage}{\selectlanguage}
\makeatletter
\DeclareRobustCommand{\selectlanguage}[1]{%
  \@ifundefined{alias@\string#1}
    {\ORIGselectlanguage{#1}}
    {\begingroup\edef\x{\endgroup
       \noexpand\ORIGselectlanguage{\@nameuse{alias@#1}}}\x}%
}
\newcommand{\definelanguagealias}[2]{%
  \@namedef{alias@#1}{#2}%
}
\makeatother
\definelanguagealias{en}{english}
\definelanguagealias{English}{english}
\usepackage{graphicx}
\usepackage{amsmath}
\usepackage{mathtools}
\usepackage{amsfonts}
\usepackage{amssymb,bbding}
\usepackage{bm}
\usepackage{color}
\usepackage[percent]{overpic}
\usepackage{soul}
\usepackage{amssymb}
\usepackage{wasysym}
\usepackage{dsfont}
\usepackage{float}
\usepackage{braket}
\usepackage{subfiles}
\usepackage{epstopdf}

\usepackage[normalem]{ulem}

\usepackage{tikz}
\usetikzlibrary{positioning}

\usepackage{blkarray}
\usepackage{multirow}

\usepackage{hyperref}
\hypersetup{
    bookmarks=false,         
    unicode=false,          
    pdftoolbar=false,        
    pdfmenubar=true,        
    pdffitwindow=false,     
    pdfstartview={FitH},    
    pdftitle={},    
    pdfauthor={Authors},     
    pdfsubject={},   
    pdfcreator={},   
    pdfproducer={}, 
    pdfnewwindow=true,      
    colorlinks=true,       
    linkcolor=black,          
    citecolor=blue,        
    filecolor=magenta,      
    urlcolor=blue           
}

\renewcommand{\vec}[1]{{\bf #1}}

\usepackage{soul}
\usepackage{relsize}
\usepackage{bbold} 
\DeclareMathOperator{\Tr}{Tr}
\renewcommand{\Re}{\mathfrak{Re}}

\begin{document}

\title{Influence matrix approach to many-body Floquet dynamics}

\author{Alessio Lerose}\thanks{These two authors contributed equally to this work}
\affiliation{Department of Theoretical Physics, University of Geneva, Quai Ernest-Ansermet 30, 1205 Geneva, Switzerland}
\author{Michael Sonner}\thanks{These two authors contributed equally to this work}
\affiliation{Department of Theoretical Physics, University of Geneva, Quai Ernest-Ansermet 30, 1205 Geneva, Switzerland}
\author{Dmitry A. Abanin}
\affiliation{Department of Theoretical Physics, University of Geneva, Quai Ernest-Ansermet 30, 1205 Geneva, Switzerland}
\date{\today}

\begin{abstract}

Recent experimental and theoretical works made much progress towards understanding non-equilibrium phenomena in thermalizing  systems, which act as thermal baths for their small subsystems, and many-body localized ones, which fail to do so.
The description of time evolution in many-body systems is generally challenging due to the dynamical generation of quantum entanglement.
In this work, we introduce an approach to study quantum many-body dynamics, inspired by the Feynman-Vernon influence functional.
Focusing on a family of interacting, Floquet spin chains, we consider a Keldysh path-integral description of the dynamics. The central object in our approach is the \textit{influence matrix} (IM), which describes the effect of the system on the dynamics of a local subsystem. For translationally invariant models, we formulate a self-consistency equation for the influence matrix.
%
For certain special values of the model parameters, we obtain an exact solution which represents a \textit{perfect dephaser} (PD). Physically, a PD corresponds to a many-body system that acts as a perfectly Markovian bath on itself: at each period, it measures every spin. For the models considered here, we establish that PD points include dual-unitary circuits investigated in recent works. In the vicinity of PD points, the system is not perfectly Markovian, but rather acts as a bath with a short memory time. In this case, we demonstrate that the self-consistency equation can be solved using matrix-product states (MPS) methods, as the IM \textit{temporal entanglement} is low. A combination of analytical insights and MPS computations allows us to characterize the structure of the influence matrix in terms of an effective ``statistical-mechanics" description. We finally illustrate the predictive power of this description by analytically computing how quickly an embedded impurity spin thermalizes. The influence matrix approach formulated here provides an intuitive view of the quantum many-body dynamics problem, opening a path to constructing models of thermalizing dynamics that are solvable or can be efficiently treated by MPS-based methods, and to further characterizing quantum ergodicity or lack thereof.

\end{abstract}

\maketitle

%
%
%
%
%
%
%
%
%
%
%
%
%
%

\section{Introduction}

Describing non-equilibrium quantum matter and harnessing it for quantum technology is one of the central challenges in modern physics. The problem of highly non-equilibrium dynamics of many-body systems, both isolated and open, has been attracting intense experimental and theoretical interest over the past years~\cite{BlochColdAtoms,JorgReview,Polkovnikov-rev}. Ergodic isolated systems are believed to thermalize as a result of their quantum evolution; qualitatively, such a system can act as an efficient thermal bath for its sufficiently small subsystems. Recent breakthroughs identified classes of systems that do not reach thermal equilibrium~\cite{Huse-rev, AbaninRMP, Altman-rev, ALET2018498} and therefore may exhibit new phenomena not envisioned within the framework of statistical mechanics.

Floquet systems, where the Hamiltonian is periodically varied in time, play a special role in the family of non-equilibrium systems, thanks to their natural experimental realizations. Although periodic driving sequences have been utilized in nuclear magnetic resonance for decades~\cite{Abragam}, recent works revealed a range of new surprising phenomena in Floquet systems. In particular, it was shown that many-body Floquet systems may exhibit new topological properties~\cite{Rudner13}. Many-body localization in Floquet systems can protect them from heating~\mbox{\cite{Ponte15, Lazarides15, Abanin20161}}, enabling new non-equilibrium states of matter with remarkable properties not attainable in thermal equilibrium~\cite{Khemani16, Else16, Nathan17, Choi16DTC, Zhang16DTC}.


\begin{figure*}[t]
\begin{tabular}{cc}
\centering
\includegraphics[height=0.27\textwidth]{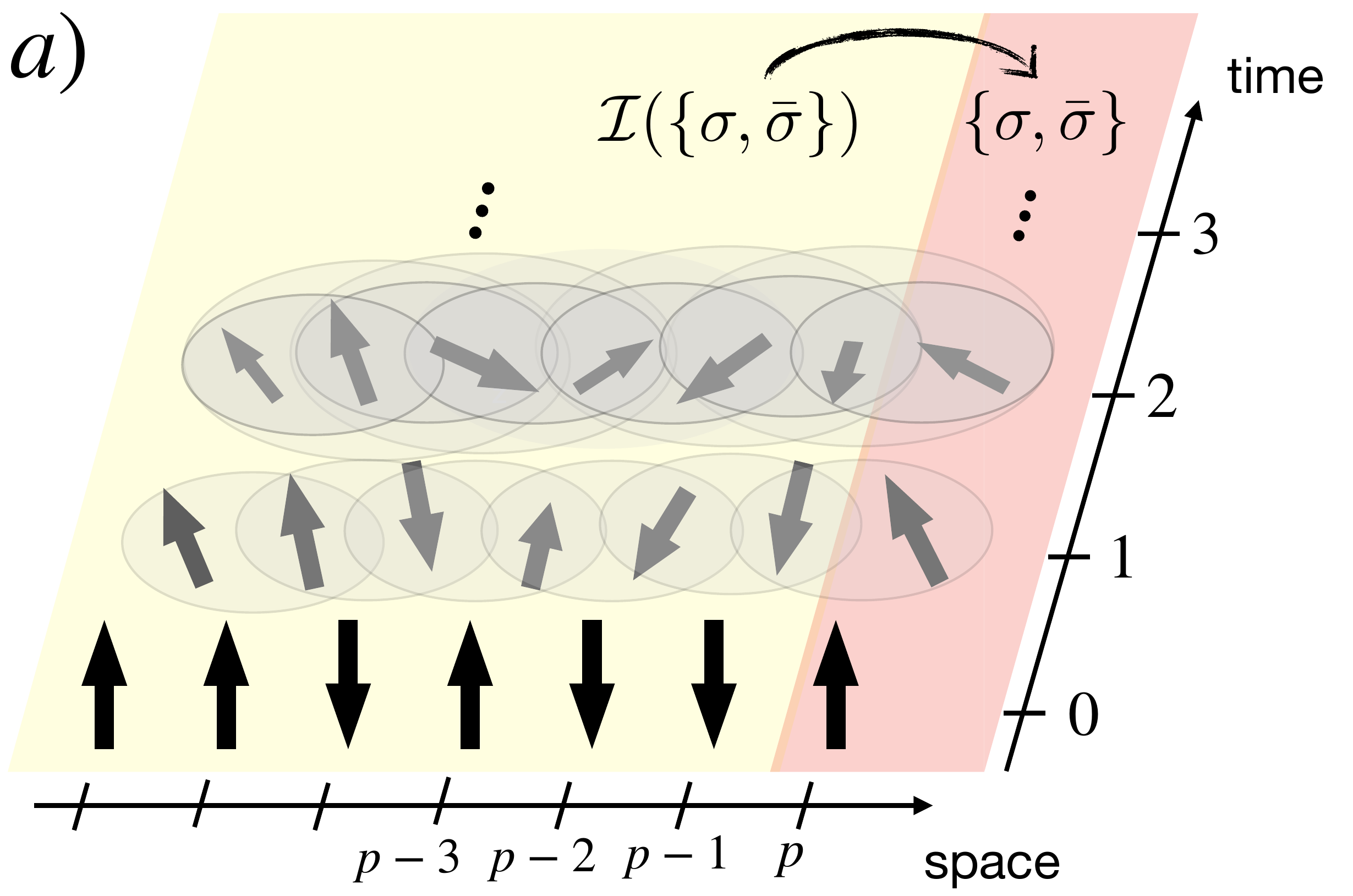} & \hspace{0.2cm}
\includegraphics[height=0.27\textwidth]{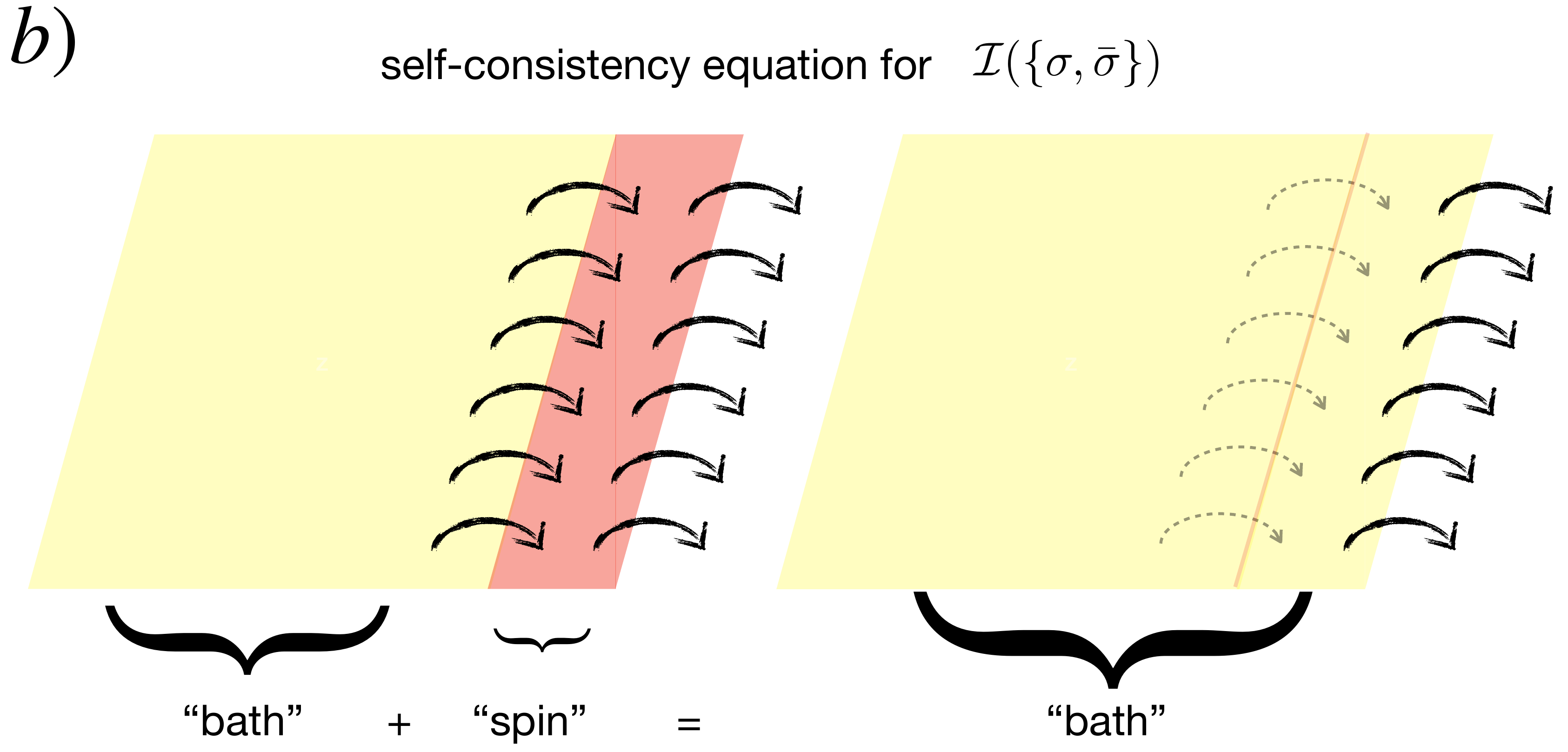}
\end{tabular}
\caption{%
a) Cartoon illustration of the influence matrix approach to many-body Floquet dynamics.
The many-body quantum state of a spin system becomes increasingly entangled due to periodic local interactions and kicks.
The influence matrix $\mathcal{I}$, defined in Eq.~\eqref{eq_IMdef} below, describes the dynamical effects of all spins $k<p$  on a spin $p$, in terms of a path integral  weight affecting the trajectories $\{\sigma,\bar\sigma\}$ of spin $p$ forward and backward in time.
b) In translationally-invariant one-dimensional systems, the influence matrix is the same for all $p$, which, as illustrated, leads to a self-consistency equation for it.
}
\label{fig_0}
\end{figure*}

The central difficulty in describing dynamics of many-body systems that do thermalize stems from the rapid generation of quantum entanglement. Initially simple, non-entangled states, quickly develop non-local correlations; faithfully describing a time-evolved state requires, in general, a number of parameters which grows exponentially with the evolution time. Various efficient numerical methods based on tensor networks have been introduced
~\cite{tebd,Banuls09,Schollwock_ReviewMPSTimeEvolution,Verstraete_ReviewMPSTangentSpace}. Examples of tractable interacting Floquet models, both integrable and thermalizing, have been found and are being actively investigated~\cite{GritsevFloquet, ChalkerPRX, Guhr1, Bertini2019,garratt2020}.

The goal of this paper is to formulate what we call the \textit{influence matrix} approach to quantum many-body Floquet dynamics. This approach can be viewed as an 
extension of the celebrated Feynman-Vernon influence functional approach~\cite{FeynmanVernon} to interacting Floquet systems. In the original formulation, Feynman and Vernon developed a path-integral description of a quantum-mechanical system coupled to a bath. Typically, the bath consists of physical degrees of freedom that are of different nature than those composing the system itself,
such as in the case of two-level systems coupled to a bath of harmonic oscillators~\cite{LeggettRMP,TEMPO} or particle reservoirs~\cite{IFnanodevices}. In our case, in contrast, the bath and the system will be composed of the same physical constituents.

We will consider quantum systems on a lattice, and will be interested in the dynamics of a finite subsystem, treating its complement as a bath. The effect of this bath on the subsystem can be described by the influence functional.
Although here we focus  on Floquet systems, generalizations to Hamiltonian systems appear possible. More specifically, we will study a class of kicked Floquet systems, which can be viewed as many-body extensions of the celebrated quantum rotor model~\cite{FIshman}. In this case, the influence functional becomes discrete  and we will therefore refer to it as the influence matrix (IM).

The setup and the key idea of the approach are summarized in Fig.~\ref{fig_0}. We consider a
one-dimensional system of quantum spins, or qudits, $\sigma_k$ (with local Hilbert space
dimension $q$). We choose spin $p$, and treat all degrees of freedom to the left 
as a bath, as illustrated in Fig.~\ref{fig_0}a.
The starting point of our analysis is the Keldysh path-integral formulation of time evolution. In this formalism, forward and backward trajectories of spins arise; for spin $k$ they are denoted by $\sigma_{k}^{\tau}, \bar\sigma_{k}^{\tau}$, $\tau=0,1,2,\dots$. To capture the effect of the
bath on spin $p$, we trace out the bath degrees of freedom, which gives rise to the influence matrix $\mathcal{I}(\{\sigma_{p}^{\tau},
\bar\sigma_{p}^{\tau}\})$.
It enters as an additional weight into the path integral describing the evolution of the reduced density matrix of spin $p$. In general, this influence matrix is non-local in time.

For translationally invariant systems, spin $p$, subject to the IM
$\mathcal{I}(\{\sigma_{p}^{\tau}, \bar\sigma_{p}^{\tau}\})$, should produce
exactly the same IM for its right neighbor. Using this observation, we will formulate a self-consistency
equation for the IM, 
pictorially illustrated in Fig. \ref{fig_0}b. This equation is a key ingredient of our approach.

The knowledge of the IM allows to determine the local dynamical properties of the system, including all 
temporal correlation functions. The IM naturally incorporates the initial state, and allows averaging over ensembles of initial states. The self-consistent IM describes how a system acts as a bath on {itself}.
This approach thus gives access to detailed information regarding thermalization time scales and memory time of the bath. As discussed below, it also allows one to analyze the effect of the bath on an impurity spin. In this sense, the self-consistent IM provides a more complete characterization of a many-body system as a bath, compared to spectral properties (such as presence or absence of level repulsion~\cite{ALET2018498}), and the statistics of matrix elements studied both in the context of the eigenstate thermalization~\cite{Polkovnikov-rev} and many-body localization~\cite{Serbyn-16}.

While in general the self-consistency equation for the IM is complicated, 
it admits exact solutions in special cases. For kicked Ising models (see, e.g., Ref.~\cite{Guhr1} and references therein) with certain parameter values, we find that the infinite-temperature IM can be obtained exactly. This solution describes a bath that is a {\it perfect dephaser} (PD), that is, its effect on a spin at each step of the Floquet evolution is to exactly cancel off-diagonal matrix elements of its reduced density matrix. Phrased differently, the bath measures every spin of the system at each time step. Interestingly, for the case of kicked Ising models, the perfect dephaser class coincides with models that can be recast as dual-unitary circuits introduced  recently~\cite{Guhr1,Bertini2019}. 
In addition, as discussed below, the PD form of the IM arises upon ensemble-averaging over random realizations of the model.

Perfect dephaser systems serve as remarkably simple quantum baths: when coupled to an impurity spin, the system would act on it as an exactly Markovian bath, with a relaxation rate that depends on the coupling strength.
This is remarkable, as Markovianity is usually an approximation 
which requires the internal dynamics of the bath to be much faster than the quantum system that it measures.
Thus, the IM approach allows one to identify quantum systems which act as Markovian baths. 

At the mathematical level, the IM approach bears a similarity to a
tensor-network numerical method introduced by Ba{\~n}uls {\it et
al.}~\cite{Banuls09} for modelling  dynamics of Hamiltonian systems. Building on
the previous insights~\cite{Banuls09}, we apply a matrix-product state (MPS)
ansatz to construct the IM away from the PD points. This tool allows us to shed light on the more general structure of the IM in ergodic Floquet systems.

At the PD points, the IM, viewed as a ``wave function" in the space of single-spin trajectories, is effectively non-entangled. We further find that away from PD points,  the
IM ``wave function" exhibits slow growth of temporal entanglement, which allows us to analyze the system's dynamics at longer times than those accessible
via exact diagonalization. This observation provides a tool for identifying regimes of thermalizing Floquet dynamics that are amenable to efficient
MPS-based methods.

To characterize the structure of IMs in ergodic systems detuned away from PD points, we adopt a statistical-mechanics-like description, viewing ``quantum" intervals of a spin trajectory (i.e., the intervals where the forward and the backward path differ, $\sigma\neq \bar\sigma$) as ``particles". We study the weights of these particles and their interactions, demonstrating that in thermalizing systems 
they decay with their temporal distance. We further use this insight to predict how the system thermalizes a slower impurity spin, finding a good agreement with numerical simulations.

The influence matrix approach has several additional attractive features. Perhaps most importantly, it provides a direct, physically intuitive way to describe a many-body system as a quantum bath for its constituent parts, giving access to relevant time scales and various correlation functions, including the Loschmidt echo (see below).
Furthermore, it allows one to describe dynamics for different ensembles of initial states and in the thermodynamic limit. As mentioned above, the IM approach admits certain exact solutions that are likely not limited to PDs which will be our focus here.

The rest of the paper is organized as follows. In Section \ref{sec:IM} we introduce the IM approach, and formulate the self-consistency equation for the IM. In Section~\ref{sec:PD} we will discuss the cases where the IM can be found exactly, and has a perfect dephaser form. Further, Section~\ref{sec:pert} is dedicated to dynamics away from PD points; we introduce and justify the use of MPS-based methods, develop an analytical characterization of the IM structure, and discuss implications for dynamics.
Finally, in Section~\ref{sec:conclusions} we will summarize our results, and provide an outlook.

%
%
%
%
%
%
%
%
%
%

\section{Influence matrix formulation}\label{sec:IM}

We will start by introducing the models to be considered. We then describe a path integral representation of the dynamics. The correlation functions are expressed in terms of a transfer matrix which acts on the space of single-spin trajectories. We discuss general properties of such transfer matrices, arguing in particular that they have a \textit{pseudoprojection} property.
 The eigenvector of the transfer matrix encodes the dynamical properties in the thermodynamic limit, and allows one to find all temporal correlation functions. We interpret the eigenvector as an influence matrix, and formulate a self-consistency equation for it.

\subsection{Model}

We consider a class of kicked one-dimensional Floquet systems. At each site of a periodic chain of length $L$, we place a spin, or qudit, with $q$ basis states, denoted by $|\sigma\rangle$. During each driving period,  a two-spin operator $\hat{{P}}_{j+1/2}$ acts on all neighboring pairs $j$, $j+1$. This operator represents an Ising-type coupling: it is diagonal in the $|\sigma_j,\sigma_{j+1}\rangle=|\sigma_j\rangle\otimes |\sigma_{j+1}\rangle$ basis, and symmetric under exchange $j \leftrightarrow j+1$ :
\begin{equation}
\hat{{P}}_{j+1/2}|\sigma_j,\sigma_{j+1}\rangle= e^{i\phi_{j+1/2}(\sigma_j,\sigma_{j+1})} |\sigma_j,\sigma_{j+1}\rangle
\end{equation}
with $\phi(\sigma,\sigma')=\phi(\sigma',\sigma)$.
This is followed by unitary single-spin operators $\hat{W}_j$ (``kicks"). 
The combination of the two steps gives the following Floquet operator (evolution operator over one driving period):
\begin{equation}\label{eq:floquet_operator}
\hat{F}= \hat W \hat P = \prod_j \hat{W}_j \prod _j \hat{P}_{j+1/2} \, .
\end{equation}
A graphical representation of the corresponding Floquet evolution is provided in Fig.~\ref{fig_1}. Here we choose to focus on this class of models for clarity of presentation. The following discussion, however, can be extended to more general Floquet systems, where a shallow quantum circuit is applied periodically in time. For $q=2$, 
this model reduces to the kicked Ising model (KIM), which we will frequently use for illustration purposes.

\subsection{Dual transfer matrix}
\label{sec:pathintegral}

Next, we use a discrete path-integral representation of the dynamics of the system. An amplitude for the evolution over $t$ periods, from an initial product state $|\{ \sigma_j^0 \} \rangle=\otimes _{j} |\sigma_j^0\rangle $ at $\tau=0$ to a state $|\{ \sigma_j^t \}\rangle$ at time $\tau=t$, is given by
$$
A_{\{\sigma_j^0 \}\to \{ \sigma_j^t \}}=\langle \{\sigma_j^t \} |  \hat{F}^t|\{\sigma_j^0 \}\rangle.
$$
This amplitude can be expressed by a path integral:
\begin{equation}\label{eq:amplitude}
A_{\{\sigma_j^0 \}\to \{ \sigma_j^t \}}=\sum_{\{\sigma_j^\tau \}}
\prod_{\tau=0}^{t-1} \prod_j e^{i\phi_{j+1/2}(\sigma_j^{\tau},\sigma_{j+1}^{\tau})}
[W_j]_{\sigma_j^{\tau+1}\sigma_j^\tau}
\end{equation}
where we have defined
$[W_j]_{\sigma_j^{\tau+1}\sigma_j^\tau} = \langle \sigma_j^{\tau+1}| \hat{W}_j|\sigma_j^\tau\rangle$.
The sum runs over all possible trajectories of $L$ spins, with fixed initial and final configurations.

\begin{figure*}[t]
\centering
\includegraphics[width=0.75\textwidth]{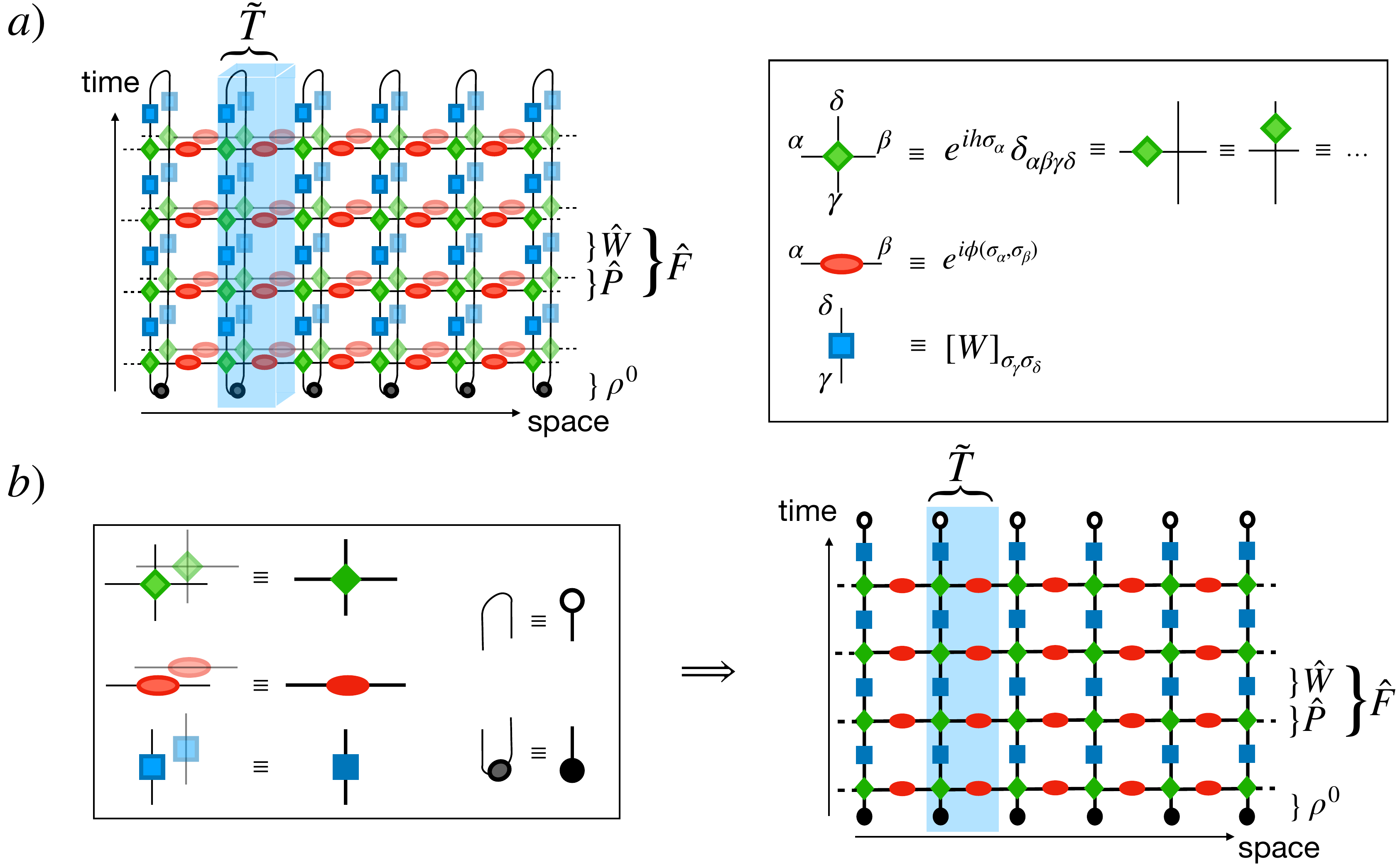}

\caption{%
a) Graphical circuit representation of the Floquet dynamics of the model in Eq.~\eqref{eq:floquet_operator}.
The periodic unitary time evolution of the initial unentangled density matrix $ \rho^0$ is pictured by a Keldysh closed-time contour made up of a forward and a backward branch. The red-ellipse and blue-square tensors represent $\hat P$ and $\hat W$ unitary operators, respectively. The black tensors denote the initial density matrices of individual spins. For later convenience, we have separated possible phases due to on-site magnetic fields, $e^{i h \sigma}$, denoted by green-diamond tensors;
the four-index Kroenecker delta tensor $\delta_{\alpha\beta\gamma\delta}$, denoted by a cross, allows to ``slide'' a green diamond leftward/rightward or upward/downward, indicating that the field $h$ can be thought of as being part of either $\hat W$ or $\hat P$. The dual transfer matrix $\tilde{T}$ defined in Eq.~\eqref{eq:Ttilde} is highlighted by a blue shaded region. 
b) In the spirit of the Keldysh formalism, the folded picture is obtained by constructing the composite, $q^2$-dimensional Hilbert spaces of the forward and backward spins at equal times, spanned by the states $\{ \ket{\sigma_j^\tau,\bar \sigma_j^\tau} \}$. The local operators acting on this folded space are given by the tensor product of each forward-branch operator $\hat O = \hat W_j, \hat P_{j+1/2}$ and its conjugate $\hat O^*$ acting on the backward branch, as illustrated in the left panel.
Accordingly, the initial density matrices $[\rho^0_j]_{\sigma_j^0,\bar \sigma_j^0}$ are reshaped into vectors $[\rho^0_j]_{(\sigma_j^0,\bar \sigma_j^0)}$, and the final-time contractions (traces) $\delta_{\sigma_j^t,\bar \sigma_j^t}$ are reshaped into vectors $\delta_{(\sigma_j^t,\bar \sigma_j^t)}$, 
denoted by black and white single-leg tensors, respectively; in either case, summation over $\sigma_j^0,\bar \sigma_j^0$ or $\sigma_j^t,\bar \sigma_j^t$ is implied by tensor contraction.
}
\label{fig_1}
\end{figure*}

To describe the evolution of the density matrix, we will employ a Keldysh-type formalism. To that end, let us introduce a superoperator $R$ which describes the evolution of the system's density matrix (DM). It is the tensor product of $A$ and its conjugate $A^*$, so that its matrix elements are given by
\begin{equation}\label{eq:R}
R_{\sigma_1^0 \bar\sigma_1^0...\sigma_L^0 \bar\sigma_L^0}^{\sigma_1^t \bar\sigma_1^t...\sigma_L^t \bar\sigma_L^t}=
A_{\{\sigma_j^0 \}\to \{ \sigma_j^t \}} A^*_{\{\bar\sigma_j^0 \}\to \{ \bar\sigma_j^t \}}.
\end{equation}
Note that $\{\sigma\}$ and $\{\bar\sigma\}$ denote two independent ``paths'' or ``trajectories'', which are conventionally referred to as forward and backward in time.
For a given initial DM of the system $\left[\rho^0\right]_{\sigma_1^0 \bar\sigma_1^0...\sigma_L^0 \bar\sigma_L^0}$, the DM at time $t$ is obtained by contracting $R$ with $\rho^0$. In particular, $R$ gives the probability of the system's transition from an initial classical state to a final classical state, if we put $\sigma_j^0=\bar\sigma_j^0$ and $\sigma_j^t=\bar\sigma_j^t$ for all $j$.

Further, following Ref.~\cite{Banuls09}, we introduce a {\it dual transfer matrix}~$T_j$, which will provide a convenient alternative representation of the reduced density matrix evolution, with time and space interchanged (see Fig.~\ref{fig_1}). Its matrix elements are given by 
\begin{widetext}
\begin{equation}\label{eq:transfer_matrix}
\langle \sigma_{j+1}^1 \, \bar\sigma_{j+1}^1 \, \dots \, \bar\sigma_{j+1}^{t-1}|  \left[T_j\right]_{\sigma_j^0 \bar\sigma_j^0}^{\sigma_j^t \bar\sigma_j^t}|\sigma_j^1 \, \bar\sigma_j^1 \, \dots \, \bar\sigma_j^{t-1}\rangle=\prod_{\tau=0}^{t-1}
e^{i\phi_{j+1/2}(\sigma_j^{\tau+1},\sigma_{j+1}^{\tau+1})-i\phi_{j+1/2}(\bar\sigma_j^{\tau+1},\bar\sigma_{j+1}^{\tau+1})} \;
[ W_j]_{\sigma_j^{\tau+1}\sigma_j^\tau} [ W_j^*]_{\bar\sigma_j^{\tau+1}\bar\sigma_j^\tau} \, .
\end{equation}
We treat initial and final spin configurations as parameters. The matrix acts on the $q^{2(t-1)}$-dimensional space of single-spin forward and backward trajectories at times $1\le \tau \le t-1$.

Then, for a periodic system, Eq.~(\ref{eq:R}) can be expressed via the dual transfer matrices at different sites as follows~\footnote{For open chains, the trace is replaced by boundary vectors with uniform unit components.}:
\begin{equation}\label{eq:RviaT}
R_{\sigma_1^0 \bar\sigma_1^0...\sigma_L^0 \bar\sigma_L^0}^{\sigma_1^t \bar\sigma_1^t...\sigma_L^t \bar\sigma_L^t}={\rm Tr} \left(  \left[T_1 \right]^{\sigma_1^t \bar\sigma_1^t}_{\sigma_1^0 \bar\sigma_1^0} ... \left[T_L \right]^{\sigma_L^t \bar\sigma_L^t}_{\sigma_L^0 \bar\sigma_L^0}   \right)  \, .
\end{equation}
In the tensor-network language, this representation corresponds to contracting the network in the space direction, as discussed in Ref.~\cite{Banuls09} which introduced a new numerical method for Hamiltonian systems based on this idea.

As we will see below, this representation has several advantages. It is particularly suited for describing
ensembles of initial product states (or, more generally, { weakly entangled states}).
 For product states,
the initial DM is a tensor product of individual spins DMs  $\left[\rho_j^0\right]_{\sigma_j^0 \bar\sigma_j^0}$ :
\begin{equation}\label{eq:product_DM}
\left[\rho^0\right]_{\sigma_1^0 \bar\sigma_1^0...\sigma_L^0 \bar\sigma_L^0}=\prod_j \left[\rho_j^0\right]_{\sigma_j^0 \bar\sigma_j^0} \, .
\end{equation}
In this case, to obtain $\rho^t$ we can conveniently contract $T_j$ with the corresponding initial DM of spin $j$.
Moreover, if we are interested in the evolution of a given spin $p$, we should contract the final indices for all other spins, by putting $\sigma_j^t=\bar\sigma_j^t$, and summing over them. This new dual transfer matrix, illustrated graphically in Fig.~\ref{fig_1}, can be expressed using {Einstein} notation for tensor contraction as
\begin{equation}\label{eq:Ttilde}
\tilde{T}_j=\left[\rho_j^0\right]_{\sigma_j^0\bar\sigma_j^0} \left[T_j \right]^{\sigma_j^t \sigma_j^t}_{\sigma_j^0 \bar\sigma_j^0}\, .
\end{equation}
Then, the evolution of spin $p$'s DM is generated by the superoperator
\begin{equation}\label{eq:1spinDM}
R_{\sigma_p^0\bar\sigma_p^0}^{\sigma_p^T\bar\sigma_p^T}={\rm Tr} \left( \tilde T_1 \dots \tilde T_{p-1} \left[T_p \right]_{\sigma_p^0\bar\sigma_p^0}^{\sigma_p^T\bar\sigma_p^T} \tilde T_{p+1} \dots \tilde T_L  \right)  .
\end{equation}
It is straightforward to adapt this formalism to different boundary conditions.

Furthermore, for {\it translationally invariant} systems and initial states, all dual transfer matrices are the same,~$\tilde T_j=\tilde T$. To further simplify Eq.~(\ref{eq:1spinDM}), we first discuss properties of $\tilde{T}$. \\

\begin{figure*}
\centering
\includegraphics[height=0.4\textwidth]{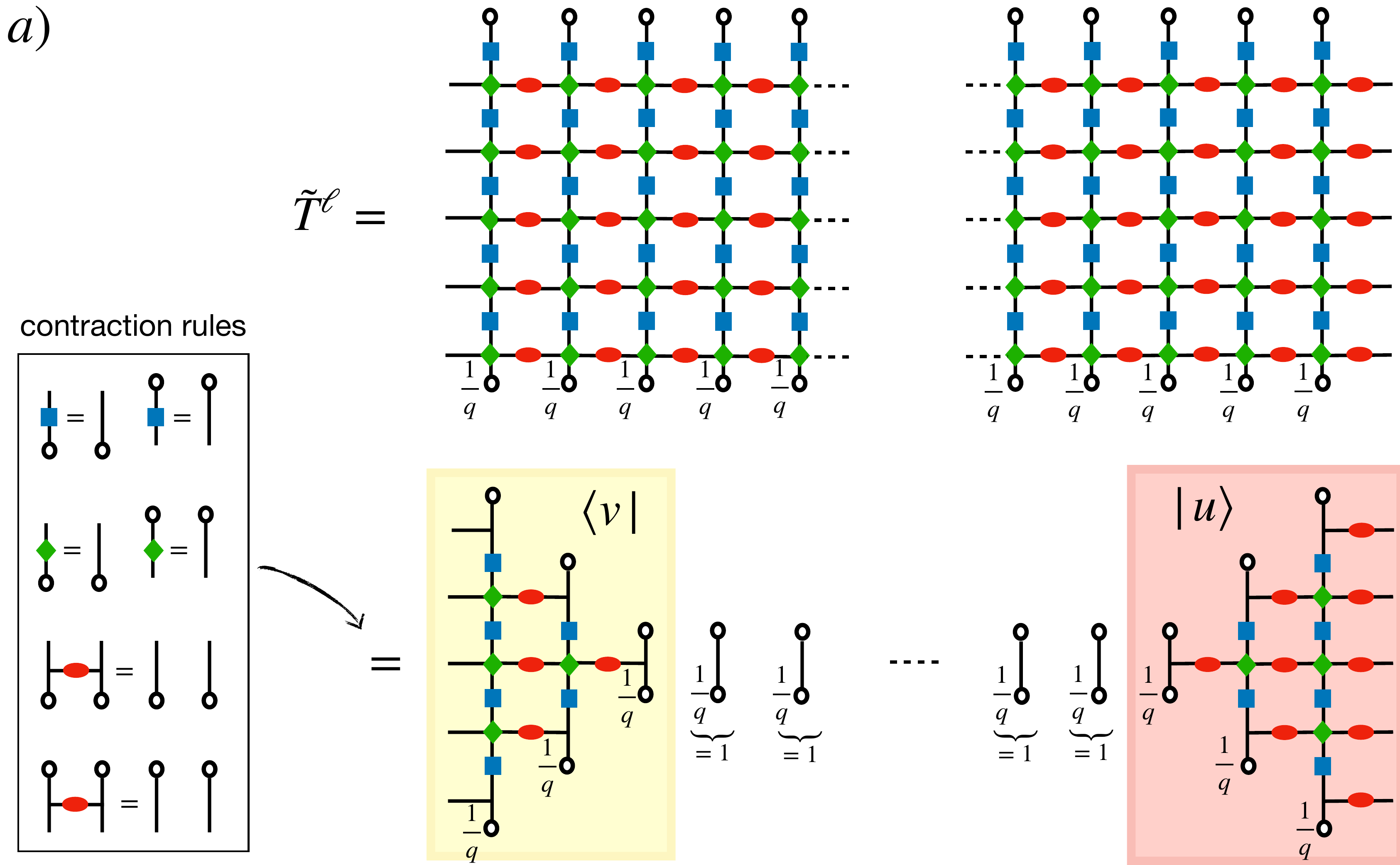}
\hspace{1cm}
\includegraphics[height=0.4\textwidth]{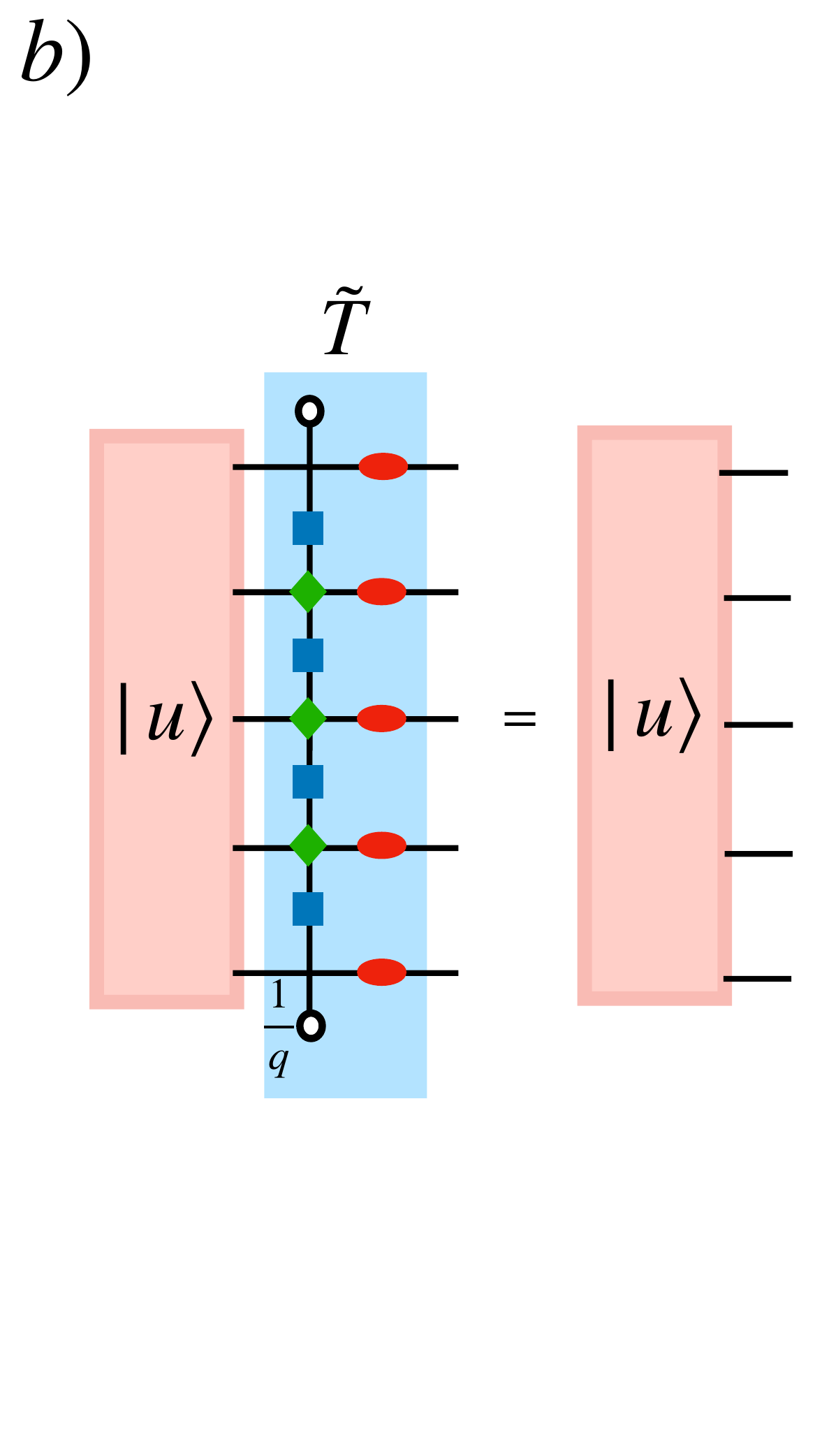}
\caption{%
a) Graphical illustration of the pseudoprojection property in Eq. \eqref{eq:decompose}.
The Keldysh path integral can be represented by a folded circuit and hence evaluated sequentially in the space direction, yielding the iteration $\tilde{T}^{\ell}$ of the dual transfer matrix (blue shaded tensor) as in Eq.~\eqref{eq:1spinDM} (cf. Fig. \ref{fig_1}).
Unitarity of time-evolution allows the contraction of the network using the rules in the framed box on the left.
Performing all possible contractions, one finds that for $\ell>t$ the input (left) and output (right) legs belong to disconnected networks, bounded by the upper and lower light cones. These two networks define vectors $\bra{v}$ and $\ket{u}$, respectively. Hence, the result can be interpreted as the decomposition in Eq. \eqref{eq:decompose}.
b) Graphical illustration of the eigenvector equation of the dual transfer matrix, which gives the self-consistency equation for the influence matrix of the system (see below).
}
\label{fig_2}
\end{figure*}

Dual transfer matrices constructed as above share the following {\it pseudoprojection} property.
Unitarity of time evolution
 implies that ${\rm Tr}(\tilde T^L)=1$ for all integer powers $L$.
 Thus, $\tilde T$ has a single non-vanishing eigenvalue $\lambda=1$. We denote by $|u\rangle$ the corresponding right
 eigenvector,  with components $ u_{\{ \sigma, \bar\sigma \}}$.
 All other eigenvalues are zero. As a result, iterations of $\tilde T$ eventually
produce a projection onto $|u\rangle$. The system size dependence is encoded in the Jordan blocks
 of $\tilde{T}$.
 Indeed, the strictly linear light-cone effect in such Floquet
models graphically illustrated in Fig. \ref{fig_2},  implies that the dimensions of Jordan blocks are upper bounded by $\ell^* = 2t$ in general (and $\ell^* = t$ for infinite-temperature initial ensembles, as in the Figure).
In fact, points located at distance $r>t$ from the considered site $p$ cannot affect the dynamics of spin $p$ before time $t$, which thus becomes size-independent for sufficiently long chains.
Hence, for $\ell>\ell^*$, we have 
\begin{equation}\label{eq:decompose}
\tilde T^\ell=|u\rangle \langle v|,
\end{equation}
where the normalization is such that $\langle v|u\rangle=1$.
From Fig.~\ref{fig_2}a, it is apparent that the left eigenvector $ \bra{v} $ differs from the right eigenvector by a multiplication by one vertical layer of the circuit, precisely the kick part of the dual transfer matrix in Eq.~\eqref{eq:transfer_matrix}:
 \begin{equation}
 v_{\{ \sigma, \bar\sigma \}}=  u_{\{ \sigma, \bar\sigma \}}
  \left[ \rho^0 \right]_{\sigma^0 \bar{\sigma}^0}
 \delta_{\sigma^t, \bar{\sigma}^t}
 \prod_{\tau=0}^{t-1}
[W]_{\sigma^{\tau+1}\sigma^\tau} [W^*]_{\bar\sigma^{\tau+1}\bar\sigma^\tau} \; .
 \end{equation}

The evolution superoperator $R$ of the density matrix of spin $p$ in an
infinite homogeneous system~\footnote{In our model, there is a strictly linear
light cone. Therefore, a system of size $L>2t$, effectively, is free of
finite-size effects, and its dynamics at times $\tau\leq t$ is equivalent to
that of an infinite system.}, Eq.~(\ref{eq:1spinDM}), can be conveniently
expressed via the eigenvector $\ket{u}$ as
\begin{equation}\label{eq:1spinDM2}
R_{\sigma_p^0\bar\sigma_p^0}^{\sigma_p^t\bar\sigma_p^t}=\langle v|\left[T_p \right]_{\sigma_p^0\bar\sigma_p^0}^{\sigma_p^t\bar\sigma_p^t}  |u\rangle
\quad =
\sum_{\{\sigma_p,\bar\sigma_p\}}
u^2_{\{\sigma_p,\bar\sigma_p\}}
\prod_{\tau=0}^{t-1}
[W_p]_{\sigma_p^{\tau+1}\sigma_p^\tau} [W_p^*]_{\bar\sigma_p^{\tau+1}\bar\sigma_p^\tau} \, .
\end{equation}
Knowledge of $R$ allows us to compute all temporal correlation functions of a single spin.
Thus, the problem of describing the dynamics of subsystems reduces to characterizing
the properties of the eigenvector of the transfer matrix $\tilde{T}$. We remind the readers that this transfer matrix depends on the ensemble of initial states, see Eq. \eqref{eq:Ttilde}.

\subsection{Self-consistency equation for the influence matrix} \label{sec:sc}

Let us take a closer look at the eigenvalue equation for $\tilde{T}$, and give
it a transparent physical interpretation. We will assume translational
invariance. We now denote the components of the eigenvector $|u\rangle$ (in the space
of trajectories of a spin on the Keldysh contour) via
\begin{equation}\label{eq:v}
|u\rangle=\sum_{\{ \sigma, \bar \sigma \}} \mathcal{I}(\{ \sigma, \bar \sigma\}) |\sigma^0 \bar \sigma^0 \dots \sigma^{t-1} \bar \sigma^{t-1}\rangle,
\end{equation}
where $\{\sigma, \bar \sigma \}$ denotes a trajectory of a spin, i.e., $\sigma^0, \bar \sigma^0, \dots , \sigma^{t-1}, \bar \sigma^{t-1}$. Then, using Eq.~(\ref{eq:transfer_matrix}), we can rewrite the eigenvalue equation as
\begin{equation}\label{eq:eig_eq}
\mathcal I(\{\sigma, \bar \sigma\})=\sum_{\{s, \bar s \}} \mathcal{I}(\{s, \bar s\})
\;  \delta_{s^t, \bar{s}^t}
\bigg( \prod_{\tau=0}^{t-1} e^{i \phi(s^{\tau},\sigma^{\tau})-i\phi(\bar s^{\tau}, \bar\sigma^{\tau}) } \,
\left[\mathcal{W} \right]_{s^\tau \bar s^\tau }^{s^{\tau+1} \bar s^{\tau+1}} \bigg)
\left[ \rho^0 \right]_{s^0 \bar{s}^0},
\end{equation}
where $\mathcal W$ is the ``folded" kick operator, which is the tensor product of $\hat{W}$ acting along the forward path $\{s\}$, and $\hat W^*$ acting along the backward path $\{\bar s\}$. 

\end{widetext}

Next, note that Eq.~(\ref{eq:eig_eq}) has a simple physical origin. We can think of the r.-h.s. as describing the propagation in time of a spin $\{s,\bar s\}$, which experiences on-site kicks $\mathcal{W}$, and which is coupled to a neighboring spin $\{\sigma,\bar\sigma\}$. Finally, the term $\mathcal I (\{s, \bar s \})$ describes the effect of all the degrees of freedom to the left of our spin on its evolution. Viewing it now as a functional of the trajectory, rather than a vector, we can interpret it as the Feynman-Vernon influence functional (see Ref. \cite{FeynmanVernon} and Appendix \ref{app_properties}), or, rather, {\it influence matrix}, given the discreteness of time.
%
In fact, the influence functional 
is obtained by tracing out the degrees of freedom to the left of a given spin $p$ (see Fig.~\ref{fig_2}): 
\begin{widetext}
\begin{equation}
\label{eq_IMdef}
\mathcal I (\{\sigma_p, \bar \sigma_p \}) =
\sum_{\substack{\{ \sigma_k^\tau, \bar \sigma_k^\tau \} \\ {k<p, \, 0\le\tau\le t}}}     \prod_{k<p}  \left[\rho_k^0\right]_{\sigma_k^0\bar \sigma_k^0}  \delta_{\sigma_k^t,\bar \sigma_k^t} \prod_{\tau=0}^{t-1}
e^{i\phi_{k+1/2}(\sigma_k^{\tau},\sigma_{k+1}^{\tau})-i\phi_{k+1/2}(\bar \sigma_k^{\tau},\bar \sigma_{k+1}^{\tau})}
\left[\mathcal{W}_k \right]_{\sigma_k^\tau \bar \sigma_k^\tau }^{\sigma_k^{\tau+1} \bar \sigma_k^{\tau+1}}.
\end{equation}
\end{widetext}
This expression describes the effect of
the environment, composed of all spins $k<p$, on the time evolution of spin $p$.
Thus, Eq.~(\ref{eq:eig_eq}) states that a spin, subject to an influence matrix $\mathcal{I}$ created by a bath to its left, creates an equal influence matrix for its neighboring spin to its right. This self-consistency equation for $\mathcal{I}$ is pictorially illustrated in Fig. \ref{fig_0}b.

In more abstract terms, the influence matrix $\mathcal I (\{ \sigma_p, \bar \sigma_p \}) $ can be regarded as the overlap between the forward and the backward propagations of the environment formed by the spins $k<p$, subject to the trajectories $\{\sigma_p\}$, $\{\bar \sigma_p\}$ of the spin $p$, respectively:
\begin{equation}
\label{eq:generalIF}
\mathcal I (\{ \sigma_p, \bar \sigma_p \}) = {\rm Tr} \Big(  U_{k<p}[\{\sigma_p^\tau \}]  \, \rho_{k<p}^0 \,  U^\dagger_{k<p}[\{\bar \sigma_p^\tau\}] \Big).
\end{equation}
 By unitarity of quantum evolution, one has \mbox{$| \mathcal I (\{\sigma, \bar \sigma \}) | \le 1$}. We will call trajectories where the forward and backward paths are
 identical $\bar \sigma_p^\tau=\sigma_p^\tau$ {\it classical trajectories} since they are
 the equivalent of classical field configurations in the Keldysh
 formalism.
 For classical trajectories, one has $ \mathcal I (\{ \sigma,  \sigma \}) =1$.
Other general properties of the influence matrix can be obtained by extending the analysis of Ref.~\cite{FeynmanVernon} to a discrete-time evolution, {and are reported in Appendix~\ref{app_properties}}.
%

Although the self-consistency equation~(\ref{eq:eig_eq}) encodes much of the complexity of a many-body system's dynamics, we will show below that there are cases when it can be solved analytically. Furthermore, we will show that the local relaxation dynamics can be naturally linked to a statistical-mechanics interpretation of the influence matrix elements $\mathcal I (\{ \sigma, \bar \sigma \})$.

Below we will be interested in averaging over the infinite-temperature ensemble.
Thus we put \mbox{$\left[\rho_j^0
\right]_{\sigma_j^0\bar\sigma_j^0}= \delta_{\sigma_j^0
\bar\sigma_j^0} / q$} .
Other initial product states will lead to different transfer
matrices. It is also possible to consider entangled initial states, at the expense
of increasing the dimensionality of the transfer matrix.

\section{Perfect dephasers}\label{sec:PD}


The process of thermalization starting from an initial product state is accompanied by the growth of entanglement between a spin and the rest of the system. In the language of the influence matrix, this corresponds to the suppression of non-classical paths, $| \mathcal I (\{ \sigma \neq \bar \sigma \}) | < 1$. The exact form of this suppression encodes the dynamics of thermalization and the decay of correlation functions in a many-body system. In general, we may expect the IM to be a  complicated functional that depends on the precise nature of the path $\{ \sigma , \bar{\sigma} \}$; parametrizing such a functional requires a number of parameters which is exponential in evolution time.

Surprisingly, 
there is a class of 
models for which 
the exact form of the IM 
is extremely simple:
it vanishes exactly for all non-classical trajectories.
%
This solution of the self-consistency equation~(\ref{eq:eig_eq}) has a
direct physical interpretation:
The environment cancels out all the interference terms.
Phrased differently, the environment completely dephases a spin at each
evolution step, erasing the off-diagonal elements of its density matrix.
Thus, we call such models {\it perfect dephasers} (PD).

Below we discuss examples of such solvable points for $q=2$ (kicked Ising model of spins $1/2$), and generalizations of KIMs to higher spins, $q>2$. We find that for 
this family of
models, the perfect dephaser points coincide with the self-dual points introduced by Akila {\it et al.}~\cite{Guhr1}, and subsequently studied in Refs.~\cite{Bertini2019,Guhr20_ExactCorrelations,Guhr_Hadamard}. Indeed, it is possible to show that dual-unitarity implies PD property, see Refs.~\cite{Bertini2019,Piroli2020} and Subsection C below. 
There, we further show how this property is reproduced by ensemble-averaging over fully random kicks and interactions.
Throughout this Section, we will consider infinite-temperature initial ensembles.

\subsection{Kicked spin-$1/2$ Ising model}

First, we will consider the case of spin-$1/2$, $q=2$. We will choose the basis $|\sigma\rangle$ to be the eigenbasis of the $z$ spin projection operator, such that $\sigma=\pm 1$. We will also employ the conventional Pauli matrix notations, $\hat\sigma_x, \hat\sigma_y, \hat\sigma_z$.

As the single-spin kick operator $\hat{W}$, we will choose a combination of a rotation around the $z$ axis followed by a rotation around the $x$ axis,  
\begin{equation}\label{eq:W_KIM}
\hat{W}=e^{i \epsilon \hat\sigma_x} e^{i h \hat\sigma_z}.
\end{equation}
Further, the two-spin term that depends on phases $\phi(\sigma,\sigma')$, reduces to the Ising interaction with a coupling strength $J$:
\begin{equation}
\label{eq:phi_KIM}
\phi(\sigma,\sigma')=J\sigma\sigma'.
\end{equation}
Thus, for the $q=2$ case, our model is equivalent to the much studied kicked Ising model (KIM), which is known to display a variety of dynamical regimes, depending on the values of parameters $h,J,\epsilon$. In particular, for $h=0$ this model becomes solvable by mapping onto free fermions via Jordan-Wigner transformation. In general, when all three parameters are non-zero, the model is non-integrable, and obeys the eigenstate thermalization hypothesis (ETH), and exhibits thermalizing dynamics~\cite{Kim_ETH}.

Next, we assume that the influence matrix has a perfect dephaser (PD) form,
\begin{equation}\label{eq:PD}
\mathcal{I}_{PD}(\{s, \bar s\})=\prod_{\tau=0}^{t-1} \delta_{s^\tau \bar s^{\tau}} \, .
\end{equation}
Let us plug this PD influence matrix (IM) into the self-consistency equation (\ref{eq:eig_eq}). We will see that for some special choices of the system's parameters $\epsilon, J$, this IM indeed solves the equation. With this form of $\mathcal I$, the summation in the r.-h.s. of Eq.~(\ref{eq:eig_eq}) is performed only over classical trajectories, so one has to keep track just of $s^\tau$ (since $\bar s^\tau=s^\tau$). Then, the r.-h.s. can be rewritten using a one-dimensional transfer-matrix, composed from the matrix elements of $\hat{W}$ and $e^{iJ s^\tau\sigma^\tau}$. { The field $h$ drops out and its value can be arbitrary}. Then the self-consistency equation takes the following form:
\begin{equation}\label{eq:SC_rewritten}
\prod_{\tau=0}^{t-1} \delta_{\sigma^\tau\bar\sigma^\tau}=
\frac 12 (1 \;\; 1)
\Bigg(
\overset{\longleftarrow}{\prod_{\tau=0}^{t-1}} \,
B(\sigma^\tau,\bar\sigma^\tau) \,
A
\Bigg) \,
\left( \begin{array}{c}
1 \\
1
\end{array}\right)
\end{equation}
where
\begin{align} \label{eq:AB}
B(\sigma^\tau,\bar\sigma^\tau) &= \left(\begin{array}{cc}
   e^{iJ(\sigma^\tau-\bar \sigma^\tau)}   & 0 \\
    0 & e^{-iJ(\sigma^\tau-\bar\sigma^\tau)}
\end{array} \right)
\\
A &=
\left(\begin{array}{cc}
   \cos^2 \epsilon   & \sin^2\epsilon \\
    \sin^2\epsilon & \cos^2\epsilon
\end{array} \right) ,
\end{align}
and the arrow over the matrix product in Eq.~(\ref{eq:SC_rewritten}) denotes time-ordering.
The boundary vector in the r.-h.s. corresponds to the infinite-temperature averaging. For $\epsilon=\pm\pi/4$, $J=\pm\pi/4$, this equation is satisfied. To see this, note that the $A$ matrix projects onto the vector $\frac 1 {\sqrt{2}} (1\; 1)^T$, while the expectation value of the matrix $B$ on this vector
equals $\cos(2J(\sigma^\tau-\bar\sigma^\tau))=\delta_{\sigma^\tau \bar\sigma^\tau}$.
For any non-classical configuration, at least one of the factors will be zero. For classical configurations this expression gives~$1$, and therefore the self-consistency equation~(\ref{eq:SC_rewritten}) is satisfied.

\subsection{Higher spins}


We now turn to the case $q>2$, and show that extensions of the kicked Ising
model also become perfect dephasers for a suitable choice of parameters. We
identify PD points using a generalization of the approach outlined in the previous
Subsection. As a result, we obtain examples of PDs with $q>2$ which coincide
with a family of dual-unitary models recently introduced by Gutkin {\it et
al.}~\cite{Guhr_Hadamard}. We emphasize that here our goal is to
demonstrate that the IM approach allows one to construct examples of perfect
dephasers, rather than to provide a complete classification of such solvable
cases; this is left for future work.

Instead of a spin-$1/2$, we consider a clock variable $\sigma=1,2,\dots,q$, and an Ising-like spin-spin interaction, $\phi(\sigma,s)=J \sigma s$.
Let us assume the PD solution of the IM, and derive the suitable model parameters from the self-consistency equation. To that end, we can introduce the generalizations of the transfer matrices $A, B$ from the previous Subsection, Eq.~(\ref{eq:AB}), which now become $q\times q$ matrices. The self-consistency equation will be satisfied by the PD IM, provided these transfer matrices satisfy the following properties:
\begin{align}
 &\frac 1 q \sum_\alpha B_{\alpha\alpha} (\sigma^\tau,\bar\sigma^\tau)  = \delta_{\sigma^\tau,\bar\sigma^\tau} \, ,\qquad\quad  \\
 A_{\alpha\beta} &=   A_{\alpha'\beta'} \quad \forall \alpha,\beta,\alpha',\beta'=1,\dots,q.
\end{align}
These conditions indeed hold if we fix the parameter $J=\pi/2q$, and specify the single-qudit kick $\hat W$ to be of the following form, $\langle \sigma' | \hat W_q | \sigma \rangle= \frac 1 {\sqrt{q}}e^{i \epsilon \sigma \sigma' } \, e^{ih \sigma}$, with $\epsilon=\pi/2q$ {and arbitrary $h$}.
Then, exactly the same mechanism as for the KIM with $q=2$ is effective at $q>2$, yielding the PD influence matrix $\mathcal I(\{\sigma, \bar \sigma\})=\prod_{\tau=0}^{t-1} \delta_{\sigma^\tau\bar\sigma^\tau}$.


\subsection{Relation to dual-unitary and random circuits}

\label{subs_DURUC}

\begin{figure}[t]
\centering
\includegraphics[width=0.4\textwidth]{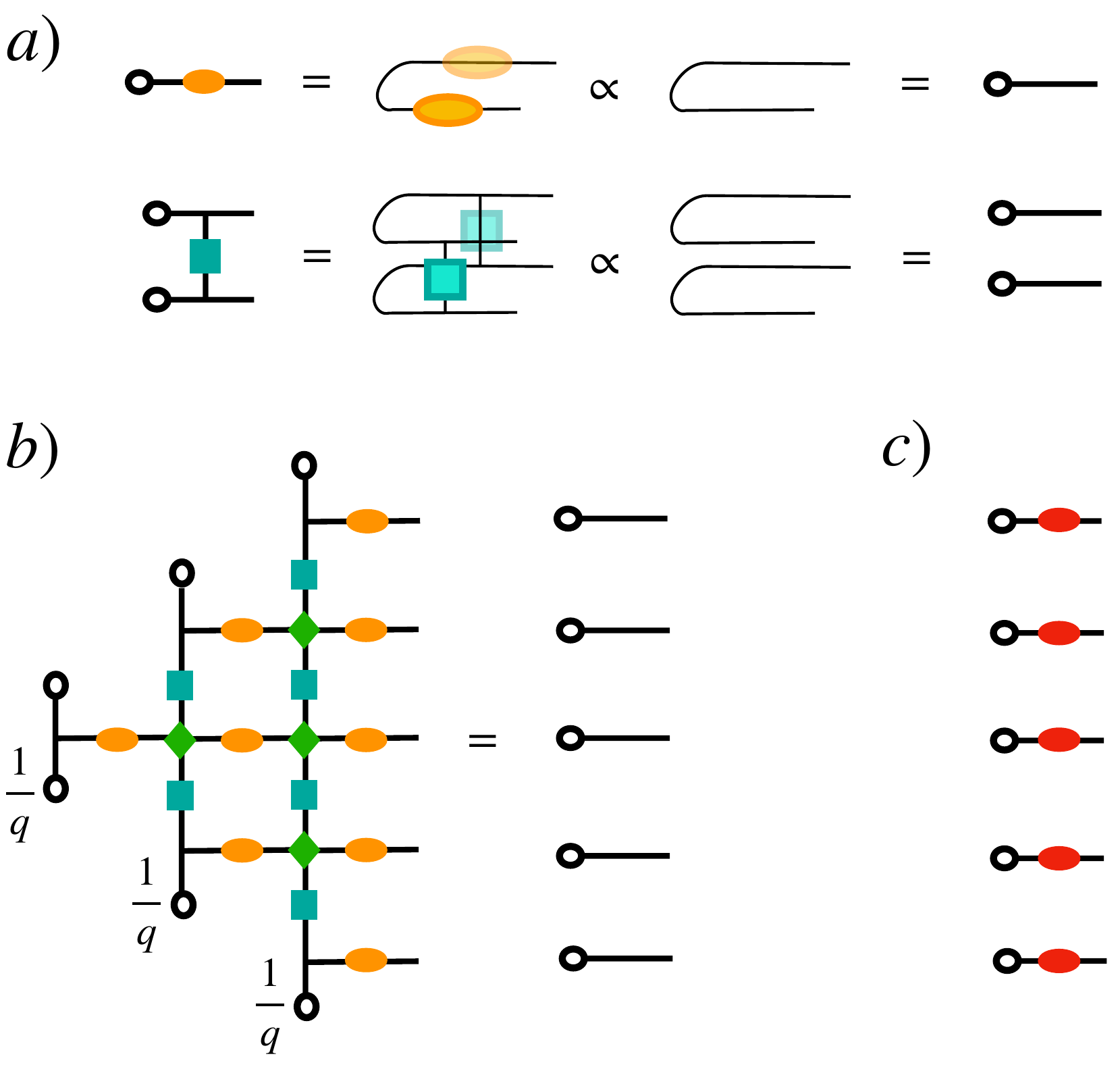}
\caption{%
a) At the dual-unitary points of the Floquet model in Eq. \eqref{eq:floquet_operator}, the depicted contractions in the space direction are allowed.
b) In this case, the network contraction proceeds within the light-cone, and the influence matrix reduces to a perfect dephaser form, i.e., a projection on classical paths. The effect on the spins on the right is that of a Markovian bath, such that coherences are cancelled at each evolution step.
c) The influence matrix acting on an impurity spin coupled to a perfect dephaser generates strictly Markovian dynamics with a dephasing strength that is tunable by changing the coupling to the spin (see Sec. \ref{subs:Markovianbath}).
}
\label{fig_3}
\end{figure}


The above examples of perfect dephasers fall into the wider class of dual-unitary circuits, as recently highlighted in Akila {\it et al.}~\cite{Guhr1} and further characterized in Refs.~\cite{Bertini2019, Rather20_CreatingDU,ClaeysPRR20,Piroli2020, Claeys20_ergodicDU}. In fact, it is possible to show that the eigenvector of the dual transfer matrix has the perfect dephaser form in Eq.~\eqref{eq:PD} whenever a circuit can be written in terms of alternating two-body unitary gates $U_{\gamma\delta,\alpha\beta}$ such that they maintain unitarity when reshaped to dual gates $\tilde{U}$ propagating along space direction, i.e., $\tilde{U}_{\beta\delta,\alpha\gamma}=U_{\gamma\delta,\alpha\beta}$~\cite{Bertini2019}. 

The proof of this statement can be graphically carried out by tensor contractions, as shown in Fig. \ref{fig_3} with tensor notations adapted to our setting. 
The parameter values of the perfect dephaser points found in previous Section allow the ``horizontal" contractions shown in panel a), which express the dual-unitary property of the gates. The result of these iterative contractions is then shown in panel b), and is nothing but the graphical representation of the PD influence matrix in Eq.~\eqref{eq:PD}.

Interestingly, the above examples of perfect dephaser circuits may be viewed as unitarily evolving systems which, in a certain sense, imitate the ensemble behavior of \textit{random} unitary circuits, recently introduced as toy models to capture structural properties of quantum dynamics generated by local interactions~\cite{NahumPRX18_OperatorSpreading}.
In fact, perfect dephasing can be enforced by ensemble averaging over spatial or spatiotemporal randomness in the interactions and kicks. Similarly to the ensemble of initial states considered above, the influence matrix approach is naturally suited to incorporate averaging over randomness in the circuit elements, provided correlations between distinct spatial points are absent; furthermore, a self-consistency equation like Eq.~\eqref{eq:eig_eq} still holds provided the distribution of randomness is translationally invariant.
In particular, the non-trivial eigenvector of the ensemble-averaged dual transfer matrix $ \mathbb{E} \big(\tilde{T}\big)$ represents the average influence of the random system on its local subsystem, and correctly generates the ensemble-averaged local observables and their temporal correlations [cf. Eq.~\eqref{eq_randomness} in Appendix~\ref{app_properties}].

To illustrate this, we show how the PD property appears in random circuit ensembles. We consider the fully random version of our model in Eq.~\eqref{eq:floquet_operator}, with random
interactions $\phi^\tau_{j+1/2}(\sigma^\tau_j,\sigma^\tau_{j+1})\in[0,2\pi)$ and random kicks
$\hat W^\tau_j \in \mathcal{U}(q)$, independently distributed in space
and time, uniformly with respect to the Haar measures.
We denote by $ \mathbb{E} (\cdot ) \equiv \prod_{j,\tau} \int d\mu_{\text{Haar}}(W^\tau_j) \int \prod_{\sigma,s} \frac{d\phi^\tau_{j+1/2}(\sigma, s)}{2\pi} \big( \cdot \big) $ the expectation value over this distribution.
The self-consistency equation~\eqref{eq:eig_eq}, which takes the form
\begin{widetext}
\begin{equation}
\label{eq_selfconsistencyrandom}
\mathcal I(\{\sigma, \bar \sigma\})=
\sum_{\{s, \bar s \}} \mathcal{I}(\{s, \bar s\})
\; \delta_{s^t \bar{s}^t}
\; \mathbb{E} \Bigg(
 \prod_{\tau=0}^{t-1} e^{i \phi^\tau_{j+1/2}(s^{\tau},\sigma^{\tau})-i\phi^\tau_{j+1/2}(\bar s^{\tau}, \bar\sigma^{\tau}) } \,
\left[ {W}^\tau_{j} \right]_{s^{\tau+1}s^\tau}
\left[ {W}^\tau_{j} \right]^*_{\bar s^{\tau+1}\bar s^\tau}
\Bigg)
\frac 1 q \delta_{s^0 \bar{s}^0},
\end{equation}
\end{widetext}
is satisfied by the perfect dephaser influence matrix $\mathcal{I}_{PD}(\{s, \bar s\})=\prod_{\tau=0}^{t-1} \delta_{s^\tau \bar s^{\tau}} $.
Since randomness is assumed uncorrelated in time, the expectation factorizes for different time steps.
Hence, the equations
\begin{equation}
\mathbb{E}\Big(  [ W^\tau_j]_{s,\sigma}  [ W^\tau_j]^*_{\bar s,\bar \sigma}         \Big) = \frac 1 q \delta_{s,\bar s}  \delta_{\sigma,\bar \sigma}
\end{equation}
and
\begin{equation}
\mathbb{E}\Big(  e^{i\phi^\tau_{j+1/2}(s,\sigma)-i\phi^\tau_{j+1/2}(\bar s,\bar \sigma)}       \Big) =  \delta_{s,\bar s}  \delta_{\sigma,\bar \sigma}
\end{equation}
lead to the solution $\mathcal{I}_{PD}$, as claimed.
We note that these equations produce contractions analogous to those satisfied by dual-unitary circuit elements as in Fig. 4a of the manuscript, although the circuit elements themselves are (almost surely in the ensemble distribution) not dual-unitary.
In this sense, dual-unitary perfect dephasers reproduce the noise of fully random models, i.e., classical white noise.

We further show that the perfect dephaser property also holds for a random \textit{Floquet} version of the circuit, with interactions and kicks uncorrelated in space but constant in time, provided the local Hilbert space is large, $q\to\infty$. We note that this model is similar to one previously considered in Ref.~\cite{ChalkerPRL18}, where random-matrix spectral correlations have been shown.
The proof of perfect dephasing follows from the mathematical properties of integration over the Haar measure of the unitary group (the so-called Weingarten calculus) in the large-$q$ limit.
Let us consider again the self-consistency equation~\eqref{eq_selfconsistencyrandom}, but now we eliminate the $\tau$-dependence of $\hat W$ and $\phi$, such that the same two random objects appear multiple times in the equation.
We want to show that the solution is again $\mathcal{I}_{PD}$.
To verify that, let us substitute $\mathcal{I}_{PD}$ into the \mbox{r.-h.s.} of Eq.~\eqref{eq_selfconsistencyrandom}. We note that in the limit $q\to\infty$ the leading contribution in $1/q$  arises from trajectories where all the $s^\tau$'s are distinct. Upon averaging, the product of the kick operators  yields $1/q^t$  (see, e.g., Ref.~\cite{Beenakker_HaarIntegration}).
%
The remaining phase term exactly equals $1$ if all $\sigma^\tau=\bar\sigma^\tau$. 
For trajectories such that $\sigma^\tau\neq\bar\sigma^\tau$ for some $\tau$'s, instead, 
the average over the random phases gives zero, unless the values of the corresponding $s^\tau$'s happen to be equal, producing the necessary phase cancellation; thus, the result is suppressed as 
$1/q$.
In other words, we have proven that
\begin{equation}
\mathcal{I}[\{\sigma,\bar\sigma\}]= \prod_{\tau=0}^{t-1} \delta_{\sigma^\tau,\bar\sigma^\tau} + \mathcal{O}\bigg(  \frac 1 q  \bigg),
\end{equation}
which is what we wanted to show.
We finally remark that it can be shown that in both the random circuit and in the random Floquet circuit model, ensemble \textit{fluctuations} around perfect dephaser average behavior are suppressed as $q\to\infty$, i.e., a random realization of the circuit is almost surely a perfect dephaser in this limit. This can be shown by evaluating $\mathbb{E}(\tilde{T}\otimes\tilde{T})$ using similar ideas as above (see also Ref.~\cite{ChalkerPRL18}).

\subsection{A many-body system that is a Markovian bath}
\label{subs:Markovianbath}

The knowledge of the influence matrix provides a complete characterization of a quantum bath and, in particular, gives a tool for describing its effect on dynamics of another quantum system coupled to it. In particular, one can analyze not just the dynamics of a translationally invariant system, but also the dynamics of an impurity immersed in the system. To illustrate this, we now study how a coupling to a perfect dephaser bath affects the evolution of a single spin.

We consider an impurity spin placed at site $p$, and coupled to its neighbors via a generic Ising coupling. We will see that the system acts on the impurity spin as a memoryless, {\it Markovian} bath. While the Markovian approximation is commonly employed in a range of problems, it normally relies on the separation of time scales between the system and the bath; in contrast, here it is {\it exact}, as a consequence of the PD property.

For simplicity, we focus on the kicked Ising model of spins $1/2$, with an impurity spin at the edge, site $p$. We choose all couplings except those involving spin $p$ to be $\epsilon=J=\pi/4$, while spin $p$ 
is coupled to spin $p-1$ by an Ising coupling with possibly different strength $\tilde J$. The kick operator $\hat W_p$ may also be different from the $\pi/2$ rotation of the other spins. Then, using the PD property, we can derive the influence matrix for the impurity spin:
\begin{equation}
\label{eq:Markovian}
\mathcal{I}(\{\sigma,\bar \sigma\})= \prod_{\tau=0}^{t-1} \cos \big(\tilde J (\sigma^\tau-\bar\sigma^\tau)\big) \, .
\end{equation}
This equation is most easily obtained graphically, noting that the last contraction on the right in Fig. \ref{fig_3}b has now non-dual-unitary gates, producing a nontrivial but factorized IM. The latter is represented in Fig. \ref{fig_3}c, where red tensors have a parameter $\tilde J$ that differs from that of the orange tensors.


The IM in Eq.~\eqref{eq:Markovian}
gives rise to a time-independent superoperator (``quantum map" or ``channel")  that evolves the local reduced single-spin density matrix,
\begin{equation}
\label{eq:Markovianmap}
 \rho_p^{\tau+1} =  \hat W_p \; \mathlarger{\mathcal{D}}\big(
 \rho_p^{\tau}
\big) \;
\hat W_p^\dagger \, ,
\end{equation}
where the dephasing superoperator $\mathlarger{\mathcal{D}}$ damps the off-diagonal entries by a factor $\cos(2\tilde J)$:
\begin{equation}
\begin{split}
\Big[\mathlarger{\mathcal{D}}\big(
 \rho_p^{\tau}
\big)\Big]_{\sigma,\sigma}
& =
\big[
 \rho_p^{\tau}
\big]_{\sigma,\sigma} \; , \\
\Big[\mathlarger{\mathcal{D}}\big(
 \rho_p^{\tau}
\big)\Big]_{\sigma,-\sigma}
&=
\cos(2\tilde J) \;
\big[
 \rho_p^{\tau}
\big]_{\sigma,-\sigma} \; .
\end{split}
\end{equation}
Unless the coupling $\tilde J$ equals $0 \mod \pi/2$, the damping is present.

The corresponding thermalization time is dictated by the leading nontrivial eigenvalue of the superoperator in Eq. \eqref{eq:Markovianmap}, which depends on both the spin's autonomous dynamics $\hat W_p$ and on the coupling $\tilde J$. 
For instance, let us consider a kick operator $\hat W_p = \exp ({i \tilde \epsilon  \hat \sigma_x})$. 
We define the temporal correlation function
\begin{equation}
\label{eq_czzdef}
\mathcal{C}_{zz}(t) =
\Big\langle
\hat F^{-t} \, \hat \sigma^z_p \, \hat F^t \, \hat \sigma^z_p
\Big\rangle_0 \, ,
\end{equation}
where $\hat F$ is the Floquet operator of the ``chain + impurity" and $\langle \cdot
\rangle_0$ denotes infinite-temperature averaging.
The polarization decay rate $\gamma_{\text{eff}}$ of the impurity spin $p$, defined by the asymptotics
\begin{equation}
\label{eq_gammamarkovian}
 - \log |\mathcal{C}_{zz}(t)| \;   \thicksim \;   \gamma_{\text{eff}} \, t,
 \end{equation}
depends on both $\tilde \epsilon$ and $\tilde J$.
In particular, if $\tilde J = \pi/4$, we have $e^{-\gamma_{ \text{eff} }} = \cos(2\tilde \epsilon)$, while if $\tilde \epsilon = \pi/4$, we have $e^{-\gamma_{ \text{eff} } }= \sqrt{\cos(2\tilde J)}$.

The above discussion shows that the effects of perfectly dephasing baths can be
computed for arbitrary times, and naturally
raises the question about the structural properties of more general quantum
ergodic baths possessing nontrivial temporal correlations. We tackle this intriguing
question below.

%
%
%
%

\section{Thermalization away from perfect dephaser points}

\label{sec:pert}


The discussion above shows that certain fine-tuned Floquet systems can act as perfect dephasing baths for themselves.
In other words, these systems can induce complete decoherence of the local degrees of freedom at each time step.
The resulting thermalization dynamics  is memoryless: i.e., it can be exactly described by a (discrete-time) Lindblad equation \cite{GKS,Lindblad}.

This property is surprising, as the Born-Markov approach to open system dynamics typically requires neglecting the bath temporal correlations \cite{BreuerPetruccioneBook}. This approximation  usually relies on a separation of timescales between the system and the bath.
While this is suitable in many physical situations involving the interaction of a ``slow" particle with distinct, ``fast" degrees of freedom (phonons, electromagnetic fields, \dots), it is generally inadequate for the local dynamics of \emph{homogeneous} extended quantum many-body systems.

The factorized form of the influence matrix in Eq. \eqref{eq:PD} of perfect dephasers, remarkably, makes their local dynamics exactly Markovian. Such property is tied with  a complete absence of \emph{temporal entanglement} in the influence matrix.
It is thus natural to investigate the structure of the influence matrix in systems detuned from  perfect dephaser points. One may expect that such systems would provide more generic examples of thermalizing quantum systems. In this Section, we shall focus on this problem.


In Sec. \ref{subs_tempentangl} we will introduce the matrix-product state (MPS) approach, which we adopt as a computational tool throughout this Section.
In Sec. \ref{subs_statmech} we unveil the underlying structure of the influence matrix of generic thermalizing systems away from PD points.
We motivate and validate a description of the IM, reminiscent of the statistical
mechanics of massive, weakly interacting particles in one dimension, where the role of particles is played by intervals of a Keldysh trajectory with $\sigma\neq\bar\sigma$.
Finally, in Sec. \ref{subs_impuritydecay}, we show that this approach can be successfully applied to compute the polarization decay rate of a slow impurity spin coupled to an ergodic chain.
For the sake of definiteness, throughout this Section we will focus on the case of the spin-$1/2$ kicked Ising model described by Eq.~\eqref{eq:floquet_operator} with the choice of operators \eqref{eq:W_KIM} and \eqref{eq:phi_KIM}.

\subsection{Low temporal entanglement and matrix-product operator approach}
\label{subs_tempentangl}

\begin{figure}[t]
\centering
\includegraphics[width=0.46\textwidth]{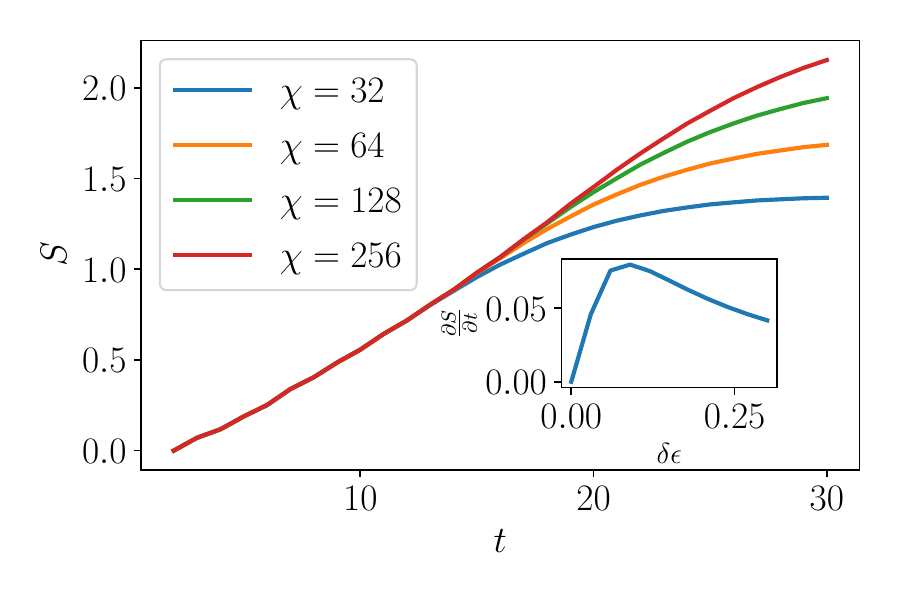}
\caption{%
Growth of the maximal (half-time) bipartite temporal entanglement entropy $S(t)$ of the
influence matrix viewed as a wave function in the folded space. It is plotted as a function of time, for a small detuning  $\delta J = \delta \epsilon = 0.06$ from the PD
point, and $h= 0.3$. Entropy growth is approximately linear in time but slow enough for the
influence matrix with $t \lesssim 20$ to be accurately captured by an MPS with a bond dimensions
$\chi\lesssim 256$.
{\it Inset:} Slope of entanglement entropy $\frac{\partial S}{\partial t}$ as a
function of detuning $\delta J = \delta \epsilon$.
As the system approaches the PD point, this slope goes to
zero. We note that the slope remains small in a broad range of detuning, enabling efficient MPS description.
}
\label{fig_entropy}
\end{figure}

It is convenient to ``fold" the backwards and forwards contour, grouping together each
spin $\sigma^\tau$ on the forward time branch and its equal-time counterpart
$\bar\sigma^\tau$ on the backward branch, to form a composite four-dimensional
local Hilbert space (cf. Fig.~\ref{fig_1}).
The influence matrix can be interpreted as a ``wavefunction" living on a chain of
those four-dimensional qudits. In this picture, the influence matrix of perfect
dephasers [Eq.~\eqref{eq:PD}] is represented by an exact product state. It is
natural to assume that for sufficiently small detuning $\delta J = \pi/4 - J$, $\delta \epsilon = \pi/4 - \epsilon$ from the PD point and arbitrary $h$, the
influence matrix can be described in the folded picture by a matrix-product state ({MPS}) with a
moderate bond dimension~$\chi$. 

We set up a code based on the TeNPy library \cite{tenpy}
which applies the dual transfer matrix $\tilde{T}$ repeatedly, starting from a product state. Due to the pseudoprojection property discussed in Sec. \ref{sec:pathintegral}, it is
sufficient to apply the dual transfer matrix $t$ times in order to obtain the
influence matrix in the thermodynamic limit. The dual transfer matrix can be expressed as a matrix-product operator (MPO)
with bond dimension 4. After each step, the influence matrix MPS is compressed
using conventional SVD truncation sweeps. Up to time {$t=2\log_2 \chi$} the compression yields no
truncation and the results from MPS match exact diagonalization {to machine precision}.

We take the convergence of the entanglement entropy $S$ of this
wavefunction upon increasing the bond dimension as a witness of the quality  of our MPS representation.
 As shown in Fig. \ref{fig_entropy},
 the MPS approach allows us to explore the properties of the IMs for larger times $t$ than those accessible via exact diagonalization,
 as $S(t)$ converges for generic values of the parameters at reasonably low bond dimensions.
The results indicate that the initial growth of $S(t)$ is approximately linear in $t$ with a slope
decreasing to zero as the PD point is approached.
Accordingly, in the following, we will use the MPS approach described here as a computational tool, with $\chi=256$.

The occurrence of low
entanglement entropy in the folded picture is supported by an intuitive
argument similar to that in Ref. \cite{muller2012tensor}: In the absence of a longitudinal field $h=0$,
the system can be described by quasiparticles which move to the left on the
forward and to the right on the backward branch. In the unfolded picture, this
leads to a strong entanglement between sites on the two branches. 
At the self dual point,
those quasiparticles can only propagate along the light cone edges
\cite{Bertini2019}, which means that in the folded picture correlations can
only exist between a forward site and its corresponding backward site.
Thus, the
entanglement entropy of the folded MPS is zero.
Detuning from the self dual
point introduces a small density of slower quasiparticles, which gives rise to a parametrically slow growth of entanglement between different folded lattice sites.


\subsection{Statistical-mechanics description of the influence matrix}
\label{subs_statmech}


\begin{figure*}[t]
\centering
\includegraphics[width=0.7\textwidth]{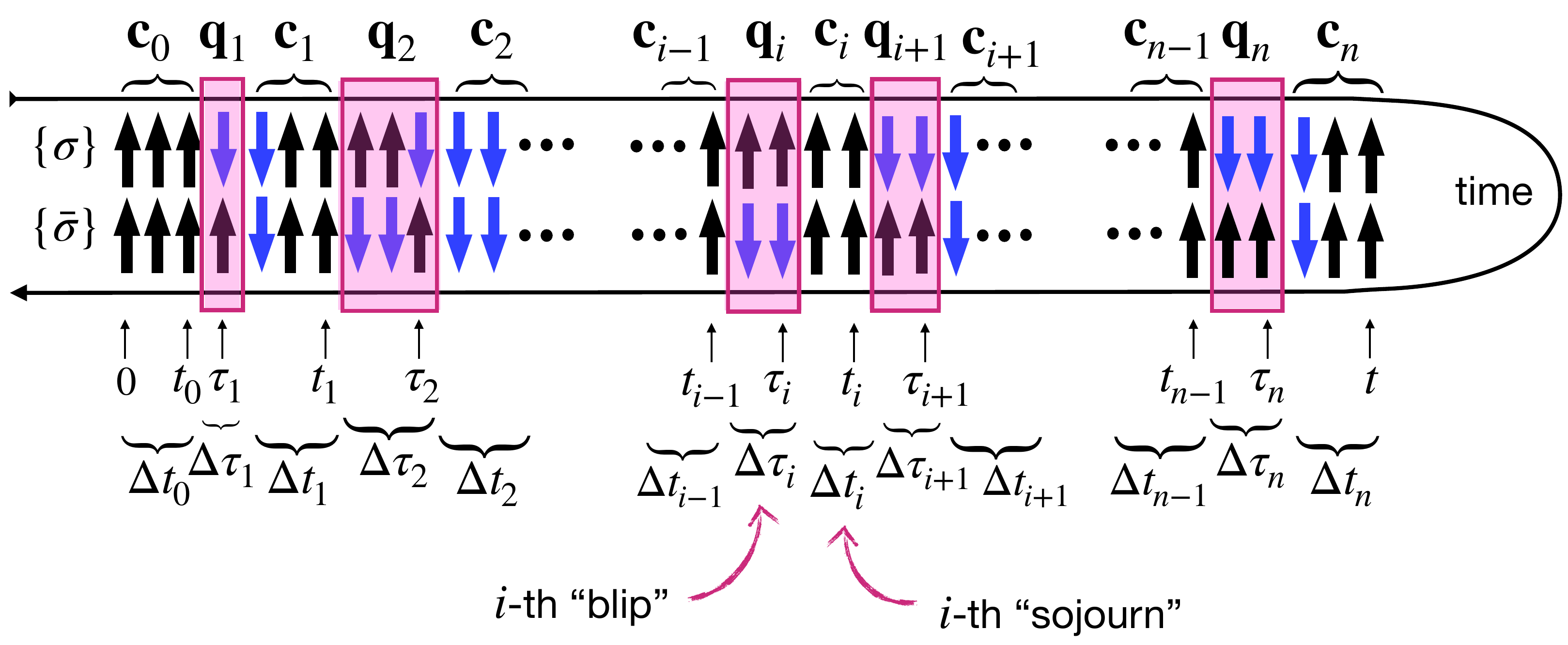}
\caption{
Illustration of the ``sojourn-blip" parametrization in Eq.~\eqref{eq_sojblip} of a generic spin trajectory on the Keldysh contour. 
}
\label{fig_keldysh}
\end{figure*}

The influence matrix is expected to develop a complex structure in generic systems away from perfect dephaser points, corresponding to the appearance of memory effects, intricate decoherence dynamics and temporal entanglement. To analyze this, it is convenient to introduce some additional formalism.
Within the previously introduced folding map (cf. Fig. \ref{fig_1}), we perform a discrete Keldysh rotation: We denote the ``classical" configurations as
\begin{equation}
\label{eq_cupdown}
\ket{\sigma^\tau \bar\sigma^\tau} =
\left\{
\begin{split}
\ket{\uparrow\uparrow} & \equiv \ket{c_\uparrow} \\
\ket{\downarrow\downarrow} & \equiv \ket{c_\downarrow}
\end{split}
\, ,
\right.
\end{equation}
and the  ``quantum" configurations as
\begin{equation}
\label{eq_qupdown}
\ket{\sigma^\tau \bar\sigma^\tau} =
\left\{
\begin{split}
\ket{\uparrow\downarrow} & \equiv \ket{q_\uparrow} \\
\ket{\downarrow\uparrow} & \equiv \ket{q_\downarrow}
\end{split}
\, .
\right.
\end{equation}
We further introduce the states
\begin{equation}
\ket{c_\pm} = \ket{c_\uparrow} \pm \ket{c_\downarrow} , \qquad
\ket{q_\pm} = \ket{q_\uparrow} \pm \ket{q_\downarrow} .
\end{equation}
Note that these four basis states may be seen as the vectorization (via the folding map) of the four basis operators $\hat{\mathbb{1}},\hat\sigma_z, \hat\sigma_x,i\hat\sigma_y$, respectively.
With this notation, the initial infinite-temperature density matrices are represented as
\begin{equation}
\frac 1 2 \mathbb{1} \rightarrow \frac 1 2 \ket{c_+}
\end{equation}
and the perfect dephaser influence matrix assumes the following simple form:
\begin{equation}
\ket{\mathcal{I}_{PD}} = 
\bigotimes_{\tau=1}^{t-1} \ket{c_+}=
\ket{ 
c_+ c_+ \dots c_+
}.
\end{equation}

General properties of the influence matrix (see Appendix \ref{app_properties} for details) dictate that $|\mathcal{I}(\{\sigma,\bar\sigma\})|\le 1$,
and $\mathcal{I}(\{\sigma,\bar\sigma\}) = 1$ for all classical trajectories, i.e.,
\begin{equation}
\mathcal{I}(c_{\uparrow,\downarrow} c_{\uparrow,\downarrow} \dots c_{\uparrow,\downarrow}) = 1
\quad \text{ or }
\quad \mathcal{I}(\{\sigma=\bar\sigma\}) = 1.
\end{equation}
More generally, as discussed in the Appendix [see in particular Eq.~\eqref{eq_generalIM}], the value of $\mathcal{I}$ on mixed classical/quantum trajectories depends only on the configuration extending between the first and the last quantum site \footnote{This is true for infinite-temperature ensembles. For general initial states, however, the influence matrix may depend nontrivially on the entire trajectory up to the last quantum state.}
\begin{equation}
\mathcal{I}\big({c_{\uparrow,\downarrow}\dots c_{\uparrow,\downarrow} (q_{\alpha_{\tau}}  \dots q_{\alpha_{\tau'}} ) c_{\uparrow,\downarrow} \dots c_{\uparrow,\downarrow}} \big)   = \mathcal{I} (q_{\alpha_{\tau}}  \dots q_{\alpha_{\tau'}} ).
\end{equation}
This property motivates us to view classical Keldysh paths as ``vacuum", and to interpret the quantum excursions $(\sigma^\tau\bar \sigma^\tau)=q_{\uparrow,\downarrow}=(\uparrow\downarrow)$ or $(\downarrow\uparrow)$ of a path $\{\sigma, \bar \sigma\}$ as ``particles". In this description, the influence matrix is regarded as a \emph{complex} statistical weight for a particle configuration,
\begin{equation}
\mathcal{I}(\{\sigma,\bar\sigma\})=e^{-\mathcal{S}(\{\sigma,\bar\sigma\})} \; .
\end{equation}
where we refer to $\mathcal{S}$ as the \emph{influence action}.
Note that $\Re \, \mathcal{S} \ge 0$.
In the limit of vanishing coupling to the bath, one trivially obtains $\mathcal{S}(\{\sigma,\bar\sigma\})=0$ for all trajectories, irrespective of their particle content.
In the opposite limit of a perfect dephaser, one has $\mathcal{I}(\{\sigma,\bar\sigma\})=0$ [i.e., $\Re \, \mathcal{S}(\{\sigma,\bar\sigma\})  = \infty$] whenever a trajectory has some particles [see Eq.~\eqref{eq:PD}].
In the generic case, a particle configuration  will be penalized by a non-zero value of the action, with  $ \Re \,\mathcal{S}(\{\sigma,\bar\sigma\}) > 0$.
Since the influence matrix contains all possible information on the dynamical effects of a part of the spin chain on the spins coupled to it, it is clear that the ``interactions" encoded in the action $\mathcal{S}$ characterize the ergodicity of the quantum dynamics, or lack thereof.

Our goal in the following is to characterize the influence matrix of generic ergodic quantum systems.
To this end, we describe Keldysh trajectories in terms of alternating classical and quantum intervals.
Following the seminal work by Leggett {\it et al.}~\cite{LeggettRMP}, we refer to
classical intervals (where $\{\sigma,\bar\sigma\}=\{\sigma,\sigma\}$) as ``sojourns" and to quantum intervals (for which $\{\sigma,\bar\sigma\}=\{\sigma,-\sigma\}$) as ``blips".
We can parametrize a trajectory as an alternating sequence of $n+1$ sojourns and $n$ blips,
\begin{equation}
\label{eq_sojblip}
\{\sigma,\bar \sigma\} \leftrightarrow \big( \vec{c}_0, \vec{q}_1,  \vec{c}_1, 
\dots, \vec{q}_n, \vec{c}_n \big) \, ,
\end{equation}
with
\begin{equation}
\vec{c}_j =( \; \underset{\Delta t_j \text{ times}}{\underbrace{c_{\uparrow,\downarrow},\dots,c_{\uparrow,\downarrow}}} \; ),
\qquad
\vec{q}_i =( \; \underset{\Delta \tau_i \text{ times}}{\underbrace{q_{\uparrow,\downarrow},\dots,q_{\uparrow,\downarrow}}} \; ),
\end{equation}
and
whose durations sum up to the total trajectory time:
\begin{equation}
\label{eq_durations}
\Delta t_0 + \Delta \tau_1 + \Delta t_1 + \dots + \Delta \tau_n + \Delta t_n = t+1.
\end{equation}
This parameterization is illustrated in Fig. \ref{fig_keldysh}.
(Here we assumed for simplicity that the initial density matrix is diagonal.)

In the following, we will use a combination of analytical insights and numerical computations to show that in quantum ergodic many-body systems, {blips behave as a ``gas" of massive, short-range interacting particles}.

\subsubsection{Blip weights}

\begin{figure}[t]
\centering
\includegraphics[width=0.46\textwidth]{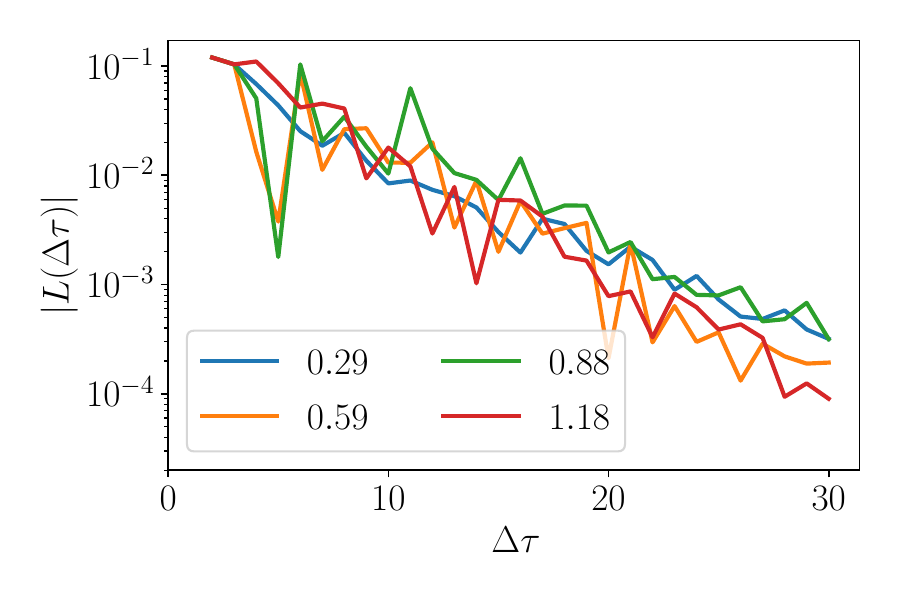}
\caption{%
Absolute value of the Loschmidt echo $L(\Delta \tau)$, equivalent to a weight of a single blip, plotted as a function of blip duration, $\Delta\tau$, near the PD point $\delta J=\delta \epsilon =
0.06$ for different values of $h$ reported in the legend. A constant blip that consists of $q_\uparrow$ spin configurations is considered.  The decay is approximately exponential, with a rate that is nearly independent of $h$, and exhibits pronounced time-dependent fluctuations around an exponential envelope.
}
\label{fig_LE}
\end{figure}


Paths without blips, $n=0$, are entirely classical, and correspond to the influence matrix $\mathcal{I}=1$.
The simplest nontrivial configurations have $n=1$, i.e., a single blip.
Their influence matrix elements, hereby called influence \emph{weights}, only depend on the internal blip structure, and not on the external sojourns.

To develop some intuition regarding the blip weights, let us first consider a constant blip,
\begin{equation}
\{ \sigma,\bar\sigma \} = \{ \sigma,-\sigma \} =
(  \underset{\Delta\tau \text{ times} }{\underbrace{q_\uparrow q_\uparrow\dots q_\uparrow}}   ) \quad
\end{equation}
(a blip with constant $q_\downarrow$ can be described in an analogous manner). In this case, it follows from the definition~\eqref{eq:generalIF} [or, more generally,~\eqref{eq_generalIM}] that the influence matrix $\mathcal{I}$ reduces to the discrete-time, infinite-temperature version of the so-called \emph{Loschmidt echo}, i.e., the overlap between two wavefunctions evolved from the same initial state with two slightly different Hamiltonians (see, e.g. Refs.~\cite{Prosen06,Pastawski17} for reviews). Denoting by a subscript $E$ the ``environment" spins at positions $0\le k<p$, we have
\begin{multline}
\label{eq_LE}
\mathcal{I}(  \underset{\Delta\tau-1 \text{ times} }{\underbrace{q_\uparrow \dots q_\uparrow}}   ) =    \Tr \bigg[  \left( \hat U_{E}^{+}  \right)^{\Delta\tau}  \frac {\mathbb{1}} {2^{p}} \left(\hat U_{E}^{- \, \dagger}  \right)^{\Delta\tau} \bigg]  \\  \equiv {L}({\Delta\tau}).
\end{multline}
The two evolution operators $ \hat U_E^{+}$ and $ \hat U_E^{-}$ differ by a classical ``field" $e^{i\phi(\sigma_{p-1},\pm 1)}=e^{\pm i J \sigma_{p-1}}$ acting on the boundary spin of the environment, due to the spin $\sigma_p$ being $\uparrow$ in the forward and $\downarrow$ in the backward evolution.

In the limit of weak coupling $J\to 0$ between spins $p-1$ and $p$, the two time evolutions in Eq.~\eqref{eq_LE} differ only slightly, and it makes sense to consider ${L}$ 
as a measure of the sensitivity of the time-evolution to small perturbations in the Floquet operator.
In fact, the continuous-time Loschmidt echo has been introduced as a quantifier of chaotic behavior in quantum systems  \cite{Peres84,Pastawski00}.
It has been found that Loschmidt echoes generically exhibit an exponential decay in ergodic systems~\cite{Prosen02_QuantumErgodicityFidelity}, while in MBL systems it displays a power-law behavior~\cite{Serbyn17}.

We are generally interested in the case of a finite, rather than weak, coupling strength $J$. Nevertheless, it is natural to
expect that in ergodic systems weights of constant blips should decay exponentially upon increasing the blip size, similar to the Loschmidt echo.
In Fig.~\ref{fig_LE} we report the results of numerical computations of the influence weight of individual blips with constant $q_\uparrow$ in the kicked Ising chain, for a range of parameters in a neighborhood of the perfect-dephaser point. In all cases, a clear exponential decay is evident, consistent with our expectations.

Furthermore, the influence weight of non-constant blips may be viewed as a generalized time-dependent Loschmidt echo of the environment, formed by the overlap of two time-evolutions subject to a classical boundary field that differs at all times. It is natural to expect  the influence weight of all blips to decay exponentially as a function of the blip duration in generic ergodic models.
To show that this exponential decay is a generic feature of long blips, we next prove that the average weight $\bar{I}(\mathbf{q}) $ of all
blips of duration $\Delta\tau$ also decays exponentially as $(\cos(2J))^{\Delta\tau}$. We start with the self-consistency equation (\ref{eq:eig_eq}) for the IM, summing over all internal configurations of a blip of length $\Delta \tau$, which yields:
\begin{multline}\label{eq:SCaverage}
  \frac{1}{2^{\Delta\tau}}\sum_{\mathbf{q} : |\mathbf{q}|=\Delta\tau} \mathcal{I}(c_\uparrow,\mathbf{q},c_\uparrow) =
  \\  \sum_{\{\sigma,\bar{\sigma}\}} \mathcal{I}(\{\sigma,\bar{\sigma}\}) \prod_{\tau=0}^{t-1} \left[\mathcal{W} \right]_{\sigma^\tau \bar \sigma^\tau}^{\sigma^{\tau+1} \bar \sigma^{\tau+1}}
  \prod_{\kappa=t_0+1}^{t_0+\Delta\tau} \cos(J(\sigma^\kappa+\bar{\sigma}^\kappa)) \, .
\end{multline}
The $\mathcal{W}$ transition amplitudes are antisymmetric with respect to the change of sign of the first sojourn. Since such a change does not affect the value of the influence matrix, the contributions of configurations with at least one blip cancel out in the average.
This leaves just the contributions from purely classical trajectories, without blips (as in Sec.~\ref{subs:Markovianbath}). For those
trajectories, the influence matrix is always $\mathcal{I}(\{\sigma,\sigma\})=1$.
Further, note that for all classical trajectories of $\sigma$ the $J$ dependent term in Eq.~(\ref{eq:SCaverage}) is the same and equal $(\cos(2J))^{\Delta\tau}$. Thus, the $\mathcal{W}$ transition amplitudes can be summed separately, which yields one. As a result, we get
\begin{equation}
  \frac{1}{2^{\Delta\tau}} \sum_{\mathbf{q} : |\mathbf{q}|=\Delta\tau} \mathcal{I}(\{c_\uparrow,\mathbf{q},c_\downarrow\}) = (\cos(2J))^{\Delta\tau} \, .
\end{equation}
We have numerically verified  this relation for various values of the model parameters, confirming that average weight of a single blip decays exponentially with its duration. Interestingly, as is evident in Fig.~\ref{fig_LE}, the weight of a blip with a fixed internal structure, exhibits fluctuations around the approximate exponential decay.

\subsubsection{Blip interactions}



\begin{figure}[t]
\centering
\includegraphics[width=0.49\textwidth]{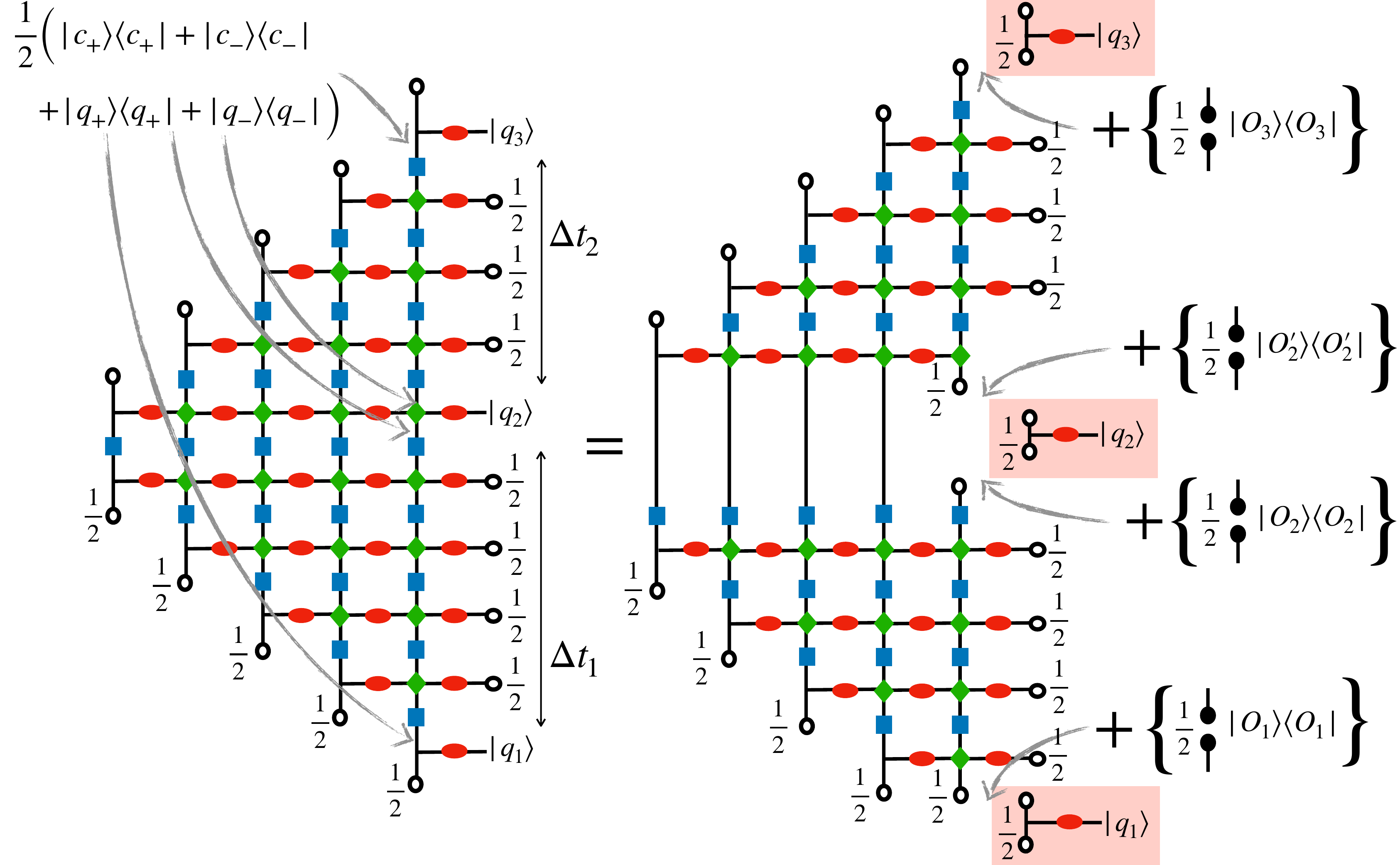}
\caption{Illustration of the decomposition of the  influence weight of a multiple-blip configuration into cluster contributions, cf. Eq. \eqref{eq_Sclusterization}.
\textit{Left-hand side:} graphical representation of the influence matrix element $ \mathcal{I}(q_1,\mathbf{c}_1,q_2,\mathbf{c}_2,q_3)$ of a configuration with $n=3$ blips of duration $\Delta\tau=1$, averaged over the intermediate sojourns $\mathbf{c}_1,\mathbf{c}_2$.
We insert spectral resolutions of the identity on each tensor leg connecting a blip with the rest of the network, as indicated by the arrows, obtaining $\mathcal{O}(4^n)$ contributions. Each of them consists of the product of $n$ disconnected networks including individual blips and the exterior network including all sojourns.
\textit{Right-hand side:}
The disconnected-blip contribution is isolated by choosing the identity state $\frac 1 2 \ket{c_+} \bra{c_+}$ in the resolution of the identity on every leg, and is highlighted by the red shading.
In this case the exterior network equals $1$, as it can be completely contracted via the rules in Fig. \ref{fig_2}a.
In contrast, for all the other contributions, the exterior network is equivalent to a temporal correlation function of traceless local operators, corresponding to states $\ket{c_-}$, $\ket{q_+}$, $\ket{q_-}$, denoted $ O_i$ here.
In ergodic dynamics, these correlations generically decay as the temporal separation $\Delta t_1$ or $\Delta t_2$ increase.}
\label{fig_blipintsketch}
\end{figure}

\begin{figure}[t]
\centering
\includegraphics[width=0.46\textwidth]{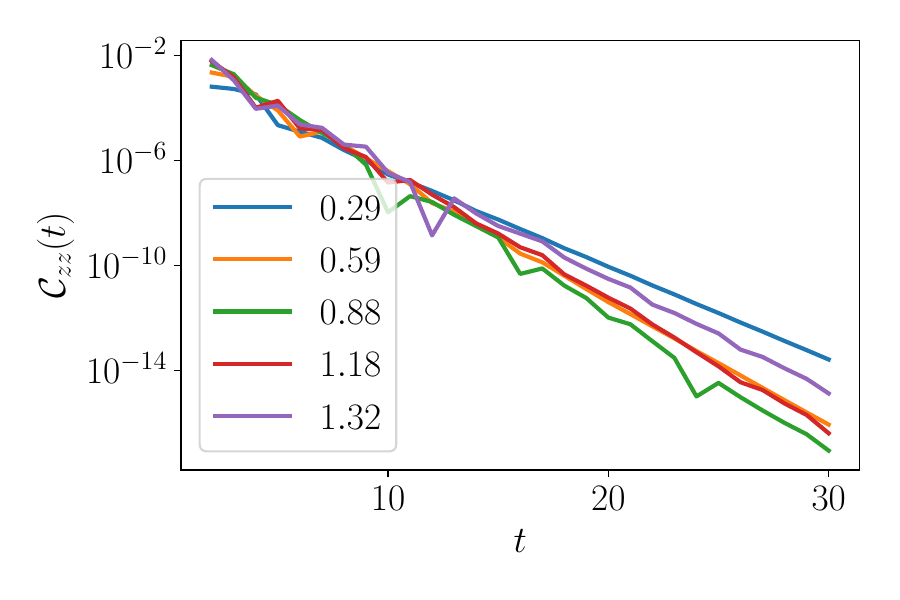}
\caption{Decay of the polarization autocorrelation $\mathcal{C}_{zz}(t)$ defined in Eq.~\eqref{eq_czzdef} vs time close to the PD point ($\delta J =
\delta \epsilon = 0.06$) for different values of $h$. The drop
is exponential with a decay rate that depends on $h$.}
\label{fig_czz}
\end{figure}

\begin{figure*}[t]
\centering
\includegraphics[width=\textwidth]{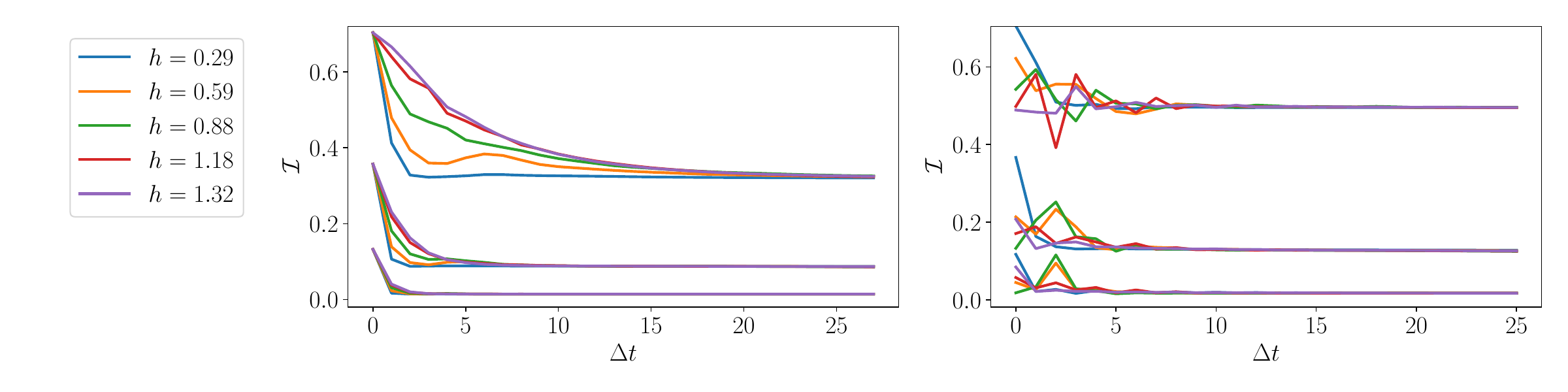} \\
\hspace{3.5cm}
\includegraphics[width=0.3\textwidth]{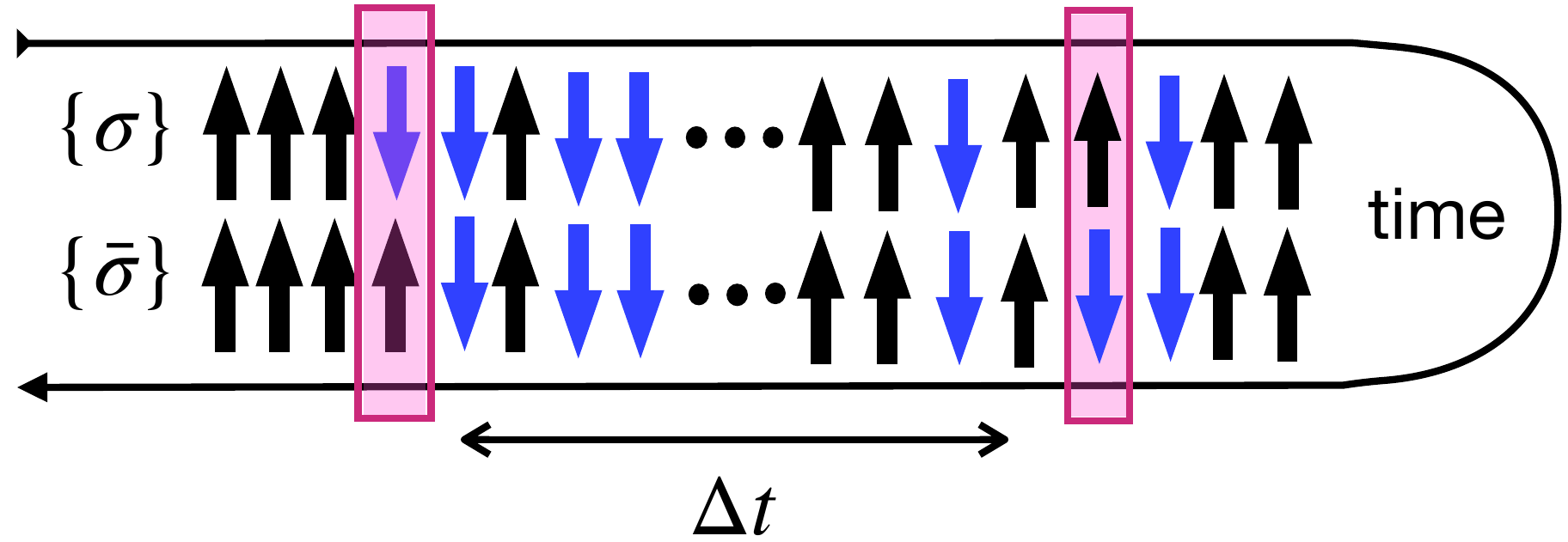}
\hspace{1.75cm}
\includegraphics[width=0.3\textwidth]{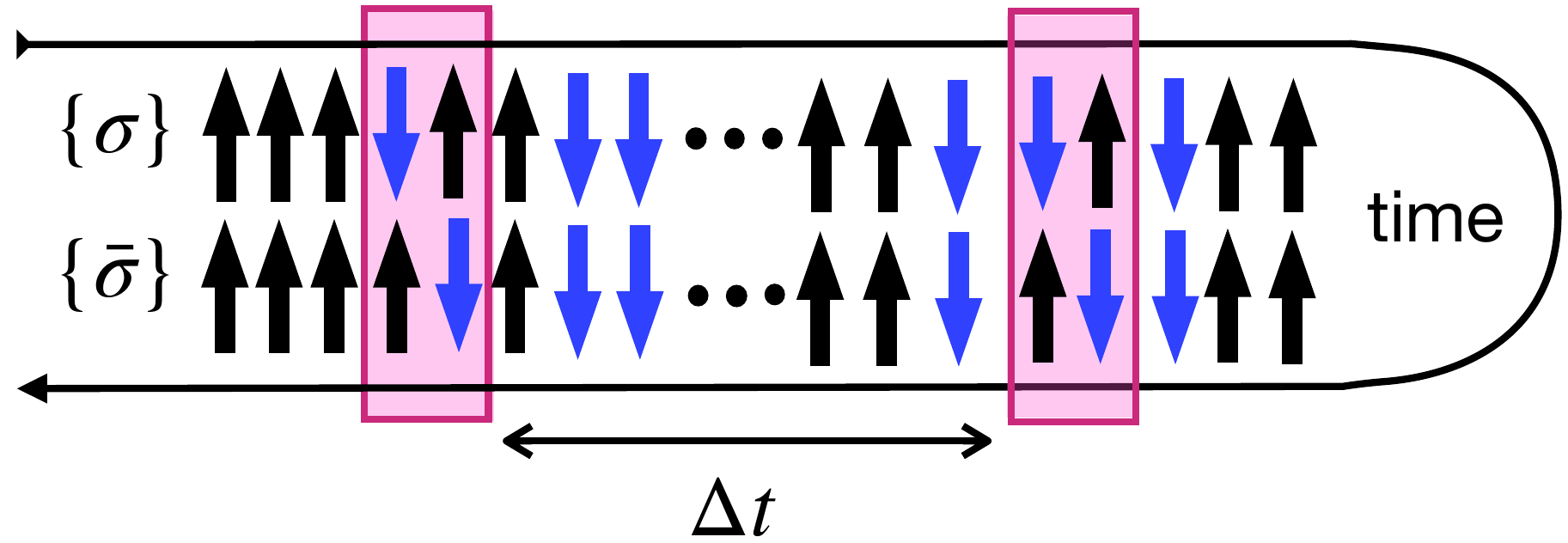}
\caption{Influence matrix elements of two blips. {\it Left (Right)}: Blips of length one
 (two) separated by sojourns of duration $\Delta t$. Detuning from the PD points,
$\delta J = \delta \epsilon=0.06,0.15,0.3$, increasing from bottom to top. Different values of $h$ are reported in the legend.
In both cases, $\mathcal{I}(\mathbf{q}_1,\mathbf{c},\mathbf{q}_2)$ decays to the product $\mathcal{I}(\mathbf{q}_1)\, \mathcal{I}(\mathbf{q}_2)$ of the values of individual blips, with $\mathcal{I}(q_{\uparrow})=\mathcal{I}(q_{\downarrow})=\cos(2 J)$ and $\mathcal{I}(q_{\uparrow}q_{\downarrow})=\mathcal{I}(q_{\downarrow}q_{\uparrow})=\cos^2(2 J)+\sin^2(2 J)\cos(2\epsilon)$.
}
\label{fig_blipint}
\end{figure*}

Within the above suggestive analogy between an IM and a statistical-mechanical complex ``action" for blips,
a long blip $\mathbf{q}$ may be pictured as a structured aggregate of  particles, the ``mass" $\mathcal{S}(\mathbf{q})$ of which increases approximately proportionally to its length.

While the weight of long blips is strongly suppressed, this need not be the case for configurations with multiple blips separated by long sojourns.
It is natural to define the \emph{interaction} influence weight  of a multiple-blip configuration as the excess influence weight  compared to the product of the influence weights of the individual blips:
\begin{multline}
e^{-\mathcal{S}_{\text{int}}( \vec{q}_1,  \vec{c}_1,  \vec{q}_2,   \vec{c}_2,   \dots, \vec{q}_n )}
= \\
e^{-\mathcal{S}( \vec{q}_1,  \vec{c}_1,  \vec{q}_2,   \vec{c}_2,   \dots, \vec{q}_n  )
+\sum_{i=1}^n \mathcal{S}( \vec{q}_i)}
\end{multline}
Note that $\mathcal{S}_{\text{int}}$  depends on the specific configuration of the sojourns $\mathbf{c}_j$ between the blips.
The question of the range of the blip interactions is intimately related to the memory time of the system.
It is expected that a weak non-Markovianity of the reduced dynamics can be translated into a fast decay of interactions.

In fact, the decay of the blip interactions may be precisely related to the temporal decay of the correlations of certain local operators, via the following argument.
Let us consider a spin trajectory with $n$ blips, $\{\ket{\mathbf{q}_i}\}_{i=1,\dots,n }$, and let us evaluate its IM element by contracting the corresponding tensor network, as shown in the example in Fig.~\ref{fig_blipintsketch} for $n=3$ and blips of duration $\Delta\tau_i=1$. (In the Figure we average over the configurations of the intermediate sojourns, although this is not necessary for the argument that follows.)
To isolate the disconnected blip contributions, we insert a spectral resolution of the identity formed by the states $\ket{c_+}$, $\ket{c_-}$, $\ket{q_+}$, $\ket{q_-}$ defined in Eqs.~\eqref{eq_cupdown}, \eqref{eq_qupdown},
on each four-dimensional tensor leg connecting a blip with the rest of the network, as indicated by the arrows in the Figure.
In this way, we have formally decomposed the IM element as a sum of $\mathcal{O}(4^n)$ contributions, each given by the product of $n+1$ disconnected network contractions: that is, $n$ ``small" single-blip networks including \mbox{$\ket{\mathbf{q}_1}$, \dots, $\ket{\mathbf{q}_n}$} respectively, and the remaining ``large" exterior network comprising all sojourns.
In this sum, the term where we choose $\frac 1 2 \ket{c_+}\bra{c_+}$ on every leg (which is explicitly represented in the Figure) yields exactly the product of the individual blip weights (red shaded), because the exterior network is equivalent to the influence weight of a classical trajectory and thus evaluates to $1$ (using the contraction rules in Fig. \ref{fig_2}a).
For all the other contributions, the large exterior network may be viewed as a (possibly complicated) temporal correlation function of local operators $\hat O_1$, \dots, $\hat O_n$, all with spatial support near the {boundary of the chain}, and with temporal support near the respective blips.
In quantum ergodic systems, such temporal correlations are expected to decay rapidly upon increasing the time separations $\Delta t_1$, \dots, $\Delta t_{n-1}$ between the operators (i.e., the duration of the intermediate sojourns).
Thus, in general, one expects the interaction strength between clusters of blips to vanish for sufficiently long sojourns in between: i.e., if $\Delta t_m \to \infty$,
\begin{multline}
\label{eq_Sclusterization}
e^{-\mathcal{S}( \vec{q}_1,  \vec{c}_1,  \vec{q}_2,   \vec{c}_2,   \dots, \vec{q}_n  )}
 \\\; \simeq  \;
e^{-\mathcal{S}( \vec{q}_1,  \vec{c}_1,    \dots, \vec{q}_m  )} \; \times \;
e^{-\mathcal{S}( \vec{q}_{m+1},  \vec{c}_{m+1},    \dots, \vec{q}_n  )}.
\end{multline}

This scenario is supported by our numerical computations.
First, we verify that temporal correlations of local operators decay upon increasing the time separation.
For the sake of illustration, we report  in Fig. \ref{fig_czz} the dynamics of the spin polarization autocorrelation $\mathcal{C}_{zz}$, defined in Eq. \eqref{eq_czzdef}. The decay occurs for all values of the parameters, as expected in generic ergodic systems.
By the above arguments, the generic decay of temporal correlations such as the ones we reported in Fig. \ref{fig_czz}  justifies the cluster approximation of the blip action.
We further verify this directly by inspecting the IM elements.
In {Fig.}
\ref{fig_blipint} we report the influence weight of a pair of blips for the
kicked Ising chain, averaged over the configurations of the intermediate
sojourns, as a function of their temporal separation, for two examples of blip
structures, and for a range of parameters in a neighborhood of the perfect-dephaser
point.
As the plots clearly show, in all cases the IM approaches a constant plateau
equal to the product of the disconnected blip weights.



\subsection{Decay rate of a slow impurity}

\label{subs_impuritydecay}





In the previous Subsection, we analyzed the influence action, finding that blips have a finite range of interactions, which is related to the relaxation time of the system.
Next, we analyze the effect of a many-body bath on a ``probe" spin embedded a spin chain, whose own dynamics is 
slower than that of the environment. In this case, thanks to the separation of time scales, the blip ``gas" is effectively dilute and can be treated as non-interacting. This will enable us to predict the relaxation rate of the impurity spin.

To make this intuition quantitative, we compute the polarization decay rate of an impurity spin  
at the edge of a kicked Ising spin chain defined by Eqs.~\eqref{eq:floquet_operator}, \eqref{eq:W_KIM}, \eqref{eq:phi_KIM}, with parameters $\epsilon=\pi/4-\delta\epsilon$, $J=\pi/4-\delta J$ detuned from a perfect dephaser point. This impurity spin differs from the other spins of the chain by the smaller strength $\tilde \epsilon$
of its kicks,
\begin{equation}
 | \tilde\epsilon| \ll \epsilon \, .
\end{equation}
Our aim is to compute the persistence of the impurity spin polarization,
\begin{equation}
P_{\sigma^0=\uparrow \, \rightarrow \, \sigma^t=\uparrow} \equiv P_{\uparrow\uparrow}(t) = \frac{1+\mathcal{C}_{zz}(t)}{2},
\end{equation}
where the autocorrelation function $\mathcal{C}_{zz}(t)$ has been previously defined in Eq.~\eqref{eq_czzdef}.

We can express this transition probability as a sum over Keldysh trajectories of the impurity spin.
Denoting the path variables by $\sigma,\bar\sigma$, we can write the above equation as follows,
\begin{widetext}
\begin{equation}\label{eq:polar_decay}
P_{\uparrow\uparrow}(t) = \sum_{
\substack{
\{\sigma^1=\uparrow,\downarrow,\; \dots, \; \sigma^{t-1}=\uparrow,\downarrow\}
\\ \{\bar\sigma^1=\uparrow,\downarrow,\; \dots, \; \bar\sigma^{t-1}=\uparrow,\downarrow\}}
}
\; \prod_{\tau=0}^{t-1}
\braket{\sigma^{\tau+1}| \exp (i \tilde\epsilon  \hat\sigma_x)  |\sigma^\tau} \braket{\bar\sigma^{\tau+1}| \exp (-i \tilde\epsilon  \hat\sigma_x) |\bar\sigma^\tau}  \; e^{ih(\sigma^\tau-\bar\sigma^\tau)}
\; \times \; \mathcal{I}(\{\sigma,\bar\sigma\}) \, .
\end{equation}
Switching to the sojourn-blip parameterization in Eq.~\eqref{eq_sojblip}, see Fig. \ref{fig_keldysh}, we can rewrite Eq.~(\ref{eq:polar_decay}) as follows:
\begin{multline}
\label{eq_pupupblipexpansion}
P_{\uparrow\uparrow}(t)
\; = \;
\sum_{n\ge 0} \;\;
(-)^n (\cos\tilde\epsilon \sin\tilde\epsilon)^{2n}
\; \times
\\
\sum_{0\le t_0 < 
\dots   < \tau_n   < t} \;\;
\sum_{
\mathbf{c}_0,\mathbf{q}_1,\mathbf{c}_1,
\dots,
\mathbf{c}_{n-1},
\mathbf{q}_n,\mathbf{c}_n
}  \;\;
\bigg(\prod_{j=0}^{n}  \widetilde{\mathcal{W}}[\mathbf{c}_j]  \bigg)
\;\;
\bigg(\prod_{i=1}^{n}  \widetilde{\mathcal{W}}[\mathbf{q}_i]  \, e^{2ih \mathcal{M}[\mathbf{q}_i]} \bigg)
\; \times \;
\mathcal{I}[\mathbf{q}_1,\mathbf{c}_1,
\dots,\mathbf{c}_{n-1},\mathbf{q}_n] \, .
\end{multline}
\end{widetext}
Now the sum over paths is arranged as a summation over the number of blips, over their positions in time, and over the internal configurations of individual sojourns and blips. In this equation, we have combined the intra-sojourn and intra-blip kick transition amplitudes into the terms
\begin{eqnarray}
\widetilde{W}[\mathbf{c}] & = (\cos^2\tilde\epsilon)^{\Delta t-1-K[\mathbf{c}]} \; (-\sin^2\tilde\epsilon)^{K[\mathbf{c}]} \\ 
\widetilde{W}[\mathbf{q}] & = (\cos^2\tilde\epsilon)^{\Delta \tau -1-K[\mathbf{q}]} \; (-\sin^2\tilde\epsilon)^{K[\mathbf{q}]}  
\end{eqnarray}
where $K[\mathbf{c}]$ ($K[\mathbf{q}]$) is the number of domain walls in the sojourn or blip configuration,
and we have defined the total magnetization a blip,
\begin{equation}
 \mathcal{M}[\mathbf{q}] = \sum_{\tau=1}^{\Delta\tau} m[q^\tau],
\quad \text{with }
m[q_\uparrow]=+1, \;
m[q_\downarrow]=-1.
\end{equation}

The above equations show that for weak kicks $|\tilde \epsilon| \ll 1$ of the impurity spin, the sum over trajectories will be dominated by terms with a low density of blips. 
In fact, the kick strength $\tilde \epsilon$ acts as a ``chemical potential" for blips.
Thus, in most of the relevant blip configurations, we may treat the blips as noninteracting, i.e., substitute
\begin{equation}
\label{eq_niba}
\mathcal{I}[\mathbf{q}_1,\mathbf{c}_1,\mathbf{q}_2,\mathbf{c}_2,
\dots,\mathbf{c}_{n-1},\mathbf{q}_n]  \; \simeq \;
\mathcal{I}[\mathbf{q}_1] \,
\mathcal{I}[\mathbf{q}_2] \,
\dots \,
\mathcal{I}[\mathbf{q}_n] \, .
\end{equation}
The success of this non-interacting blip approximation (NIBA), discussed at length by Leggett {\it et al.} in the context of a two-level system coupled to a bath of independent harmonic oscillators~\cite{LeggettRMP}, relies on the typical interblip distance being larger than the blip interaction range.

\begin{figure*}[t]
\centering
\includegraphics[width=0.46\textwidth]{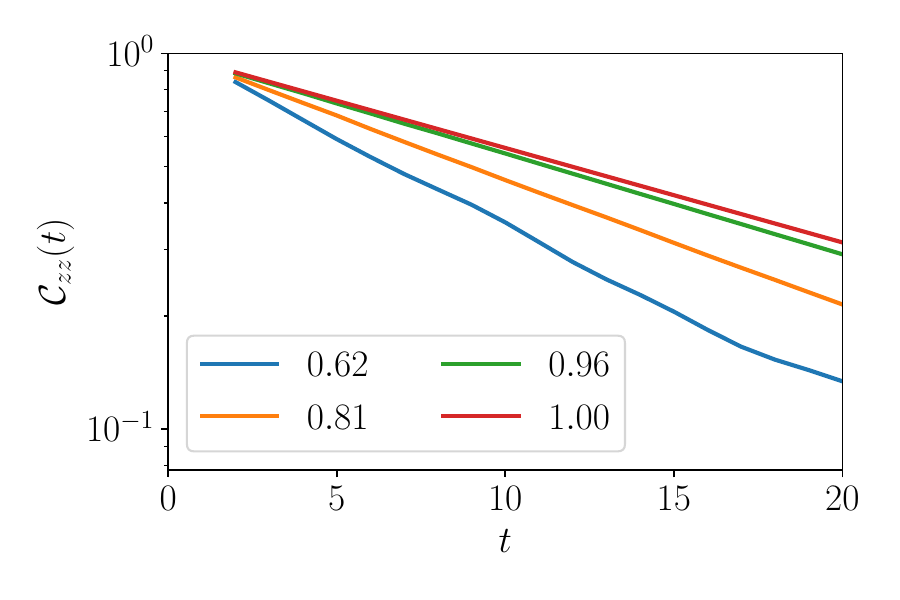}
\hspace{0.cm}
\includegraphics[width=0.46\textwidth]{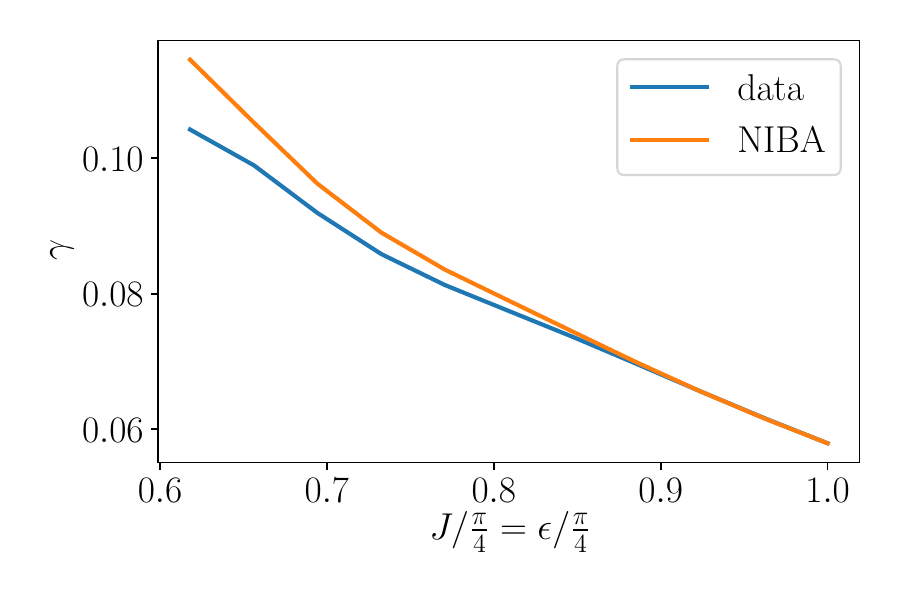}
\caption{
Left panel: dynamics of the polarization autocorrelation $\mathcal{C}_{zz}(t)$ defined in Eq. \eqref{eq_czzdef} of a slow spin, for $h=0.29;\tilde{\epsilon}=0.17$ and a range of values of $\epsilon=J$, in units of $\pi/4$ (reported in the legend).
Right panel: comparison between the numerical (blue) and the theoretically predicted (orange) decay rates vs detuning from the PD point.
The former is extracted by fitting the decay curves such as those in the left panel, while the latter is described by Eq.~\eqref{eq_nibagamma}.
}
\label{fig_slowspin}
\end{figure*}

The factorized form of the IM in Eq.~\eqref{eq_niba} allows us to analytically compute the polarization decay rate of the impurity.
After this simplification, the influence matrix in Eq.~\eqref{eq_pupupblipexpansion} does not depend on the sojourns' internal structure. The summation over all configurations of one sojourn yields
\begin{equation}
\sum_{
\mathbf{c} : |\mathbf{c}| = \Delta t
}
\widetilde{W}[\mathbf{c}]
=
2 \left( \cos(2\tilde\epsilon)  \right)^{\Delta t-1}.
\end{equation}
Note that the first and last spin are restricted to being $\sigma^0=\sigma^t=\; \uparrow$, which gives an extra overall factor of $1/4$;
moreover, the term $n=0$ has both ends fixed and gives $\frac 1 2 [1+\cos^t (2\tilde\epsilon)]$.
For blips, instead, we define the ``dressed" single-blip weight
\begin{equation}
\widetilde{\mathcal{I}}(\Delta \tau)
\equiv
\sum_{
\mathbf{q} : |\mathbf{q}| = \Delta \tau
}
\widetilde{W}[\mathbf{q}]\, e^{2ih \mathcal{M}[\mathbf{q}]}  \, \mathcal{I}[\mathbf{q}] \, .
\end{equation}
This quantity depends on how the environment suppresses individual blips, and should be considered as an input for the computation of the polarization decay.
Note that the identities $\mathcal{M}[-\mathbf{q}]=-\mathcal{M}[\mathbf{q}]$, $\mathcal{I}[-\mathbf{q}]=\mathcal{I}[\mathbf{q}]^*$
imply that $\widetilde{\mathcal{I}}(\Delta \tau)$ is real.

Since we are interested in the asymptotic decay rate of $P_{\uparrow\uparrow}(t)$, it is convenient to compute its generating function (or a discrete Laplace transform),
\begin{equation}
\bar{P}_{\uparrow\uparrow}(\lambda) = \sum_{t=0}^\infty e^{-\lambda t} P_{\uparrow\uparrow}(t).
\end{equation}
We split $t$ into intervals according to Eq.~\eqref{eq_durations},
and, for each $n$, we perform the sums over the durations of blips and sojourns.
For the blips, we define
\begin{equation}
\sum_{\Delta \tau=1}^{\infty} e^{-\lambda \Delta \tau} \widetilde{\mathcal{I}}(\Delta\tau) \equiv \bar{\mathcal{I}}(\lambda).
\end{equation}
Putting everything together, we find
\begin{widetext}
\begin{equation}
\bar{P}_{\uparrow\uparrow}(\lambda) =
\frac{1}{2}
\Bigg\{
\frac{1}{1-e^{-\lambda}} +  \frac{1}{1-    e^{-\lambda}  \big[  \cos (2 \tilde\epsilon)
- 2 \cos^2(\tilde\epsilon) \sin^2(\tilde\epsilon)  \; \bar{{\mathcal{I}}}(\lambda)   \big]}
\Bigg\}
\label{eq_laplace}
\end{equation}
\end{widetext}

The long-time behavior of the polarization is governed by the asymptotics of $\bar{P}_{\uparrow\uparrow}(\lambda)$ for small argument. In particular, an exponential decay of the correlation function $\mathcal{C}_{zz}(t)$ in Eq.~\eqref{eq_czzdef} with a rate $\gamma_{\rm eff}$ gives:
\begin{equation}
\bar{P}_{\uparrow\uparrow}(\lambda) \; \underset{\lambda\to0}{\thicksim} \; \frac{P_{\text{av}}}{\lambda} + \frac 1 {1-e^{-\gamma_{\text{eff}}}} + \mathcal{O}(\lambda),
\end{equation}
with
\begin{equation}
P_{\text{av}} = \lim_{t\to\infty} P_{\uparrow\uparrow}(t) = \frac 1 2.
\end{equation}
Comparing with the result in Eq.~\eqref{eq_laplace}, we find
\begin{equation}
\label{eq_nibagamma}
e^{-\gamma_{\text{eff}}} = \cos (2 \tilde\epsilon) - 2 \cos^2(\tilde\epsilon) \sin^2(\tilde\epsilon)  \; \bigg(\sum_{\Delta \tau}  \widetilde{\mathcal{I}}(\Delta\tau)\bigg) \, .
\end{equation}
For small $\tilde\epsilon$,  this reduces, in the lowest non-vanishing order in $\tilde{\epsilon}$, to
\begin{equation}
\label{eq_gammaeff}
\gamma_{\text{eff}}  \simeq 2 \; \tilde\epsilon^2 \;  \Big[ 1 + \sum_{\Delta \tau}  \widetilde{\mathcal{I}}(\Delta\tau)\Big].
\end{equation}
This equation expresses the decay rate of a slow impurity spin coupled to a spin chain detuned from the PD point, which is a non-Markovian bath. In the limit of a perfect dephasing bath with $|\epsilon|=|J|=\pi/4$, where the blips are completely suppressed, the above equation reduces to the Markovian dynamics, as in Eq.~\eqref{eq_gammamarkovian}. In fact, the first term in Eq. \eqref{eq_gammaeff} represents the decay rate in this limit.
This contribution is purely classical, as the environment completely suppresses interference between all pairs of quantum trajectories.
 The second term represents instead the noninteracting-blip contribution. This correction can be of either sign, due to the complex blip weights.

 Blip interactions generate higher-order corrections to~$\gamma_{\text{eff}}$. Leaving their detailed analysis for future work, we note that such corrections can be systematically taken into account via a renormalization procedure, in which one progressively includes \emph{connected} contributions of clusters of two, three, etc blips into $\widetilde{\mathcal{I}}(\Delta \tau)$. Such a scheme can be truncated as long as the series \eqref{eq_pupupblipexpansion} is dominated by terms with a low density $n/t$ of blips. For small $\tilde \epsilon$, this holds, since the blip density scales as $\mathcal{O}(\tilde \epsilon^2)$. The first correction arising from blip interactions is suppressed as $r\tilde\epsilon^2$, where $r$ is an effective parameter describing the range of two-blip interactions. Note that the value of  $r$ may depend on the structure of the blips involved and model parameters, see e.g. Fig. \ref{fig_blipint}. This will lead to a correction of the order $\mathcal{O}(r \tilde \epsilon^4)$ to the relaxation rate $\gamma_{\text{eff}}$  in Eq.~\eqref{eq_gammaeff}. For fixed $\tilde \epsilon$, this correction becomes increasingly important as one detunes the system from a perfect dephaser point, effectively enhancing the blip interaction range $r$.

We tested the predictions of Eq.~\eqref{eq_gammaeff} against our numerical computations, finding a good agreement in a broad range of model parameters.  The comparison is shown in Fig.~\ref{fig_slowspin}. In the left panel, the dynamics of the probe spin polarization autocorrelation $\mathcal{C}_{zz}(t)$ is plotted for increasing detuning $\delta \epsilon = \delta J$ of the chain from the perfect dephaser point, where the behavior is exactly exponential, $\mathcal{C}_{zz}(t) = \cos^t(2\tilde\epsilon)$. As shown, the decay of $\mathcal{C}_{zz}(t)$ remains approximately exponential even for sizeable detunings. In the right panel, we compare the measured decay rate $\gamma$ with the prediction of the NIBA in Eq.~\eqref{eq_nibagamma}. For small detunings the approximation is excellent. A small discrepancy appears for larger detunings, which we attribute to the neglected contribution of blip interactions. The latter could be accounted for via the renormalization procedure briefly sketched above. We note that we found the range of quantitative validity of the NIBA to be sensitive to the value of the integrability-breaking parameter $h$. This can be attributed to the fact that, as we saw in the previous subsection, changing $h$ tunes the range of blip interactions.


We conclude this section with some remarks.
First, the blip gas approach presented above can be straightforwardly applied to more general systems than discussed here; for example, we expect that it works similarly for the class of random models introduced in Sec.~\ref{subs_DURUC} upon detuning from the perfect dephaser limit, e.g., by decreasing the amount of randomness or the local Hilbert space dimension $q$.

Second, it is instructive to contrast this approach to the perturbation theory scheme set up in Ref.~\cite{Kos20_DUPerturbations}.
%
%
The latter work expresses space-time correlators perturbatively in the detuning from the dual-unitary points of brickwork quantum circuits, and relies on a small space-time density of perturbed gates. 
In contrast, the influence matrix approach discussed here is non-perturbative at its core.
While we  found it convenient to focus here on a neighborhood of a
dual-unitary point for conceptual clarity and clear numerical advantage, this is
not at all a crucial ingredient; for instance, an MPS representation of the IM can also be efficient in  very different regimes, e.g., when either parameter value $J$, $\epsilon$ or $h$ is small~\cite{IntegrablePaper}, or in the presence of strong disorder~\cite{Sonner20CharacterizingMBL}.
%
Similarly, the blip gas analysis presented above
is perturbative in the impurity's internal frequency scale, but not necessarily
in the detuning from PD points, which enters non-perturbatively via the blip weights and interactions. The
latter can be computed analytically or extracted numerically for general
ergodic quantum circuits.

We finally briefly comment on the difference between the discrete
spin dynamics studied above, and continuous evolution of a two-level
system coupled to a bath of harmonic oscillators~\cite{LeggettRMP}. In the``strong
decoherence limit", when blips are completely suppressed, the dynamics
in the two cases is qualitatively different. In the continuous
setting, a complete suppression of quantum interference resulting from
strong interaction with the environment freezes the spin state and
it cannot relax to equilibrium. This phenomenon is known as the
quantum Zeno effect~\cite{quantumzeno}. In the discrete case, blips are suppressed
for $\tilde{\epsilon}=\pi/4$. Then, in contrast, the spin thermalizes over just one time
step, as discussed in Sec.~\ref{subs:Markovianbath}.

\section{Summary and outlook} \label{sec:conclusions}

We have developed an approach to analyzing highly non-equilibrium dynamics of isolated many-body systems, inspired by the Feynman-Vernon influence functional formalism \cite{FeynmanVernon}. Focusing on a class of interacting Floquet models, we formulated the self-consistency equation for the influence matrix, and demonstrated that it can be analyzed using complementary analytical and numerical considerations in a whole range of model parameters. Therefore, both ``solvable" perfect dephaser baths that are Markovian, and more generic non-Markovian models can be characterized within the IM approach. As an example, we analyzed the effect of a generic many-body system on an impurity spin coupled to it.

The influence matrix provides a novel probe for characterizing quantum chaos and its absence in many-body systems. Compared to spectral probes, such as level statistics, the advantage of the IM is that it provides detailed information regarding the memory time scales and temporal correlation functions. IMs also contain the information expressed by other previously studied dynamical probes, such as the Loschmidt echo, and the statistics of matrix elements of local operators. In future work, it will be interesting to develop a connection between the properties of the IM and the eigenstate thermalization hypothesis.

We have found a family of ergodic models (in particular, in the vicinity of the perfect dephaser points), where MPS-based methods are efficient for computing the IM. This stems from viewing the IM as a ``wavefunction", whose ``temporal entanglement'' scales slowly with the evolution time.
Interestingly, conventional methods relying on the smallness of spatial
entanglement would not recognize such models as ``easy".
As a matter of fact,
this represents the key distinction from conventional time-evolution methods
based on tensor contractions sequentially in time.
While these rely on 
truncating long-distance spatial correlations built up in the course
of evolution,
the IM approach relies instead on 
 long-range correlations \textit{in time} remaining low.
This scheme could be expected to be suited to strongly chaotic quantum systems, which induce a rapid thermalization
of subsystems initially out of equilibrium, thus quickly erasing local memory of
the past.
Presumably, this different principle of efficiency leads to a new ``corner of
solvability'' in quantum many-body systems out of equilibrium, as we illustrated
with the  perfect dephaser family, where temporal entanglement is low and
spatial entanglement is high.
 This insight complements previous  works on Hamiltonian systems~\cite{Banuls09,muller2012tensor,HastingsPRAFolding}.
Interestingly, better compression schemes than conventional singular value truncations are conceivable, which could
make the influence matrix approach even more numerically efficient and broaden its range of applicability. A possible idea draws on memory kernel approximations inspired by the theory of open quantum systems~\cite{MakriMakarov94}.

Our work also suggests several promising future directions. First, it appears that PD circuits can be found in higher dimensions. Second, it would be interesting to analyze the precise relation between PD and dual-unitary circuits. As we discussed (see also Refs.~\cite{Bertini2019,Piroli2020}), dual-unitarity implies that a system is a PD, but we do not know whether the converse generally holds. Other promising generalizations of our approach include Hamiltonian systems and their dynamics at a finite, rather than infinite, temperature.

In future work, we plan to further analyze the interacting blip gas, and the effects of blip interactions on the impurity spin dynamics, as briefly discussed in the last Section. We envision that, detuning from PD points, the self-consistency equation for the IM may be solved perturbatively in the detuning. This, along with applying the IM approach to non-ergodic systems~\cite{Sonner20CharacterizingMBL}, will likely lead to a more complete characterization of non-equilibrium quantum matter.

\section{Acknowledgments}

This work was supported by the Swiss National Science Foundation. 
We thank Soonwon Choi and Ehud Altman for inspiring discussions.
Computations were performed at the University of Geneva on the ``Baobab'' HPC cluster.

\appendix

\section{ \\ General properties of influence matrices}
\label{app_properties}

In this Appendix we review the basic properties of Feynman-Vernon's influence functional, adapted from Ref.~\cite{FeynmanVernon} to our discrete-time dynamics. For simplicity, we focus on two-level systems ($q=2$); the generalization to the case of $q>2$ is straightforward.

We consider a quantum spin ($S$) interacting with an environment ($E$). In the main text discussion, $S$ is a spin at position $p$ in a spin chain, and $E$ is formed by all spins on previous positions $k<p$; here, however, for the sake of generality we keep the discussion more abstract.  We assume the spin-environment interaction to be of the form $\hat V_{\text{int}}=\hat V_S \otimes \hat V_E $. Without loss of generality, we take $\hat V_S = J \hat \sigma^z$, such that
\begin{equation}
 \hat U_{\text{int}}= e^{i(J  \hat \sigma^z \otimes \hat V_{{E}})}.
\end{equation}
The global Floquet operator has thus the form
\begin{equation}
\hat F = e^{i J \hat \sigma^z \otimes \hat  V_{\text{E}}} \; \hat U_S \otimes \hat U_E,
\end{equation}
where $\hat U_S, \hat U_E$ are the parts of the Floquet operator acting on the spin and environment separately.
We further assume that in the initial state
\begin{equation}
 \rho =  \rho_S \otimes  \rho_E,
\end{equation}
such that the spin and environment are uncorrelated.

According to the standard laws of quantum mechanics, the transition probability $P_{\sigma^0 \rightarrow \sigma^t}$ for the spin $S$ can be expressed as a discrete path integral~\cite{FeynmanHibbsBook}:
\begin{widetext}
\begin{equation}
\label{eq_transprob}
P_{\sigma^0 \rightarrow \sigma^t} = \sum_{\{\sigma^1=\uparrow,\downarrow,\; \dots, \; \sigma^{t-1}=\uparrow,\downarrow\}}  \sum_{\{\bar\sigma^1=\uparrow,\downarrow,\; \dots, \; \bar\sigma^{t-1}=\uparrow,\downarrow\}}
\; \prod_{\tau=0}^{t-1}  \braket{\sigma^{\tau+1}|\hat  U_S|\sigma^\tau} \braket{\bar\sigma^{\tau+1}|\hat  U_S|\bar\sigma^\tau} ^*
\; \times \; \mathcal{I}(\{\sigma,\bar\sigma\}) \, ,
\end{equation}
where we have collected the partial trace over the environment degrees of freedom into the \emph{influence matrix}
\begin{equation}
\label{eq_generalIM}
\begin{split}
 \mathcal{I}(\{\sigma,\bar\sigma\}) &= \Tr_E \bigg(  \hat U_E[{\{\sigma\}}] \; \hat \rho_E \;  \hat U_E^\dagger[{\{\bar\sigma\}] }\bigg) \\
 &= \Tr_E \bigg( \;
 ( \underset{\sigma^{t-1}=\pm}{\underbrace{ e^{\pm i J \hat V_{\text{E}}} }} \; \hat U_E) \; \dots  \;
 (  \underset{\sigma^{1}=\pm}{\underbrace{  e^{\pm i J \hat V_{\text{E}}} }} \; \hat U_E)
  \;\; \hat \rho_E \;\;
  ( \, \hat U_E^\dagger \; \underset{\bar\sigma^{1}=\pm}{\underbrace{    e^{\mp i J \hat V_{\text{E}}}  }} ) \; \dots \;
  ( \, \hat U_E ^\dagger \;  \underset{\bar\sigma^{t-1}=\pm}{\underbrace{   e^{\mp i J \hat V_{\text{E}}}  }} )
\bigg).
 \end{split}
\end{equation}
\end{widetext}
In this expression, the modified environment evolution operator takes the form
\begin{equation}
\hat U_E[{\{\sigma\}}]
= \overset{\longleftarrow}{ \prod_{\tau=1}^{t-1}} \;  \hat U_E[{\sigma^\tau=\pm}] 
= \overset{\longleftarrow}{ \prod_{\tau=1}^{t-1}} \; e^{\pm i J \hat V_{\text{E}}} \; \hat U_E
\end{equation}
where the arrow denotes time-ordering of the matrix product and now the c-numbers $\{ \sigma^\tau=\pm 1\}$ are given by the considered trajectory of the spin $S$. Inserting this expression into Eq.~\eqref{eq_transprob} we reconstruct the full summation over Feynman histories of the composite system.

From expression \eqref{eq_generalIM} for the influence matrix, it follows that $|\mathcal{I} | \le 1$.
In particular, it is evident that the absolute value of the influence matrix may be smaller than $1$ if the forward and backward trajectories differ. Indeed, when $\sigma^\tau=\bar\sigma^\tau$ for all $\tau$'s, the resulting ``standard" time-evolution of the density matrix preserves its trace, giving $\mathcal{I}(\{\sigma,\sigma\}) =1$.

More generally, if $\sigma^\tau=\bar\sigma^\tau$ for all $\tau>\tau_f$, the forward and backward evolution after $\tau_f$ cancel out. For this reason, we may always think of the Keldysh contour as extending up to time $+\infty$. Similarly, when the initial state is the infinite-temperature density matrix, the evolution up to $\tau_i$ cancels out if $\sigma^\tau=\bar\sigma^\tau$ for all $\tau<\tau_i$, and the Keldysh  contour can be thought as extended from time $-\infty$.

The influence matrix is generally a complicated, nonlocal functional of the spin trajectory.
As explained in Sec. \ref{subs_statmech}, we can parametrize a trajectory in terms of alternating classical ($\sigma^\tau=\bar\sigma^\tau$) and quantum ($\sigma^\tau\ne\bar\sigma^\tau$) intervals, referred to as ``sojourns" and ``blips", see Fig.~\ref{fig_keldysh}.

We finally recall two simple and useful properties of influence functionals \cite{FeynmanVernon}.
First, if the environment contains some degree of randomness, either in its Hamiltonian parameters (e.g., disorder or noisy couplings) or in its initial state (statistical mixture), the ensemble-averaged influence functional correctly describes the ensemble-averaged time-dependent observables:
\begin{equation}
 \label{eq_randomness}
\mathbb{E} \big( \mathcal{I}(\{\sigma,\bar\sigma\}) \big)
  \quad \leadsto \quad
 \mathbb{E} \big( \hat O_S(\tau) \hat O_S(\tau') \big),
  \end{equation}
where $ \mathbb{E} \big( \cdot \big)$ denotes the ensemble-averaging over randomness.
Second, if the system is coupled to multiple uncorrelated environments, the composite functional describing the simultaneous influence of all of them is the product of the individual influence functionals:
\begin{equation}
S \cup E_1 \cup \dots \cup E_n \quad \implies \quad \mathcal{I}(\{\sigma,\bar\sigma\}) = \prod_{i=1}^N \mathcal{I}_{E_i}(\{\sigma,\bar\sigma\}) \, .
\end{equation}

\bibliography{mbl}

\end{document}